\documentclass[11pt,a4paper]{article}
\usepackage{amsmath,dsfont}
\usepackage{amssymb,amsthm}
\usepackage{bm}
\usepackage{graphicx}
\usepackage{geometry}
\usepackage{mathtools}
\usepackage{acronym}
\usepackage{color}
\usepackage{enumerate}
\usepackage{appendix}
\usepackage{algorithm}
\usepackage{algpseudocode}
\usepackage{algorithmicx}
\usepackage{bm}
\usepackage{soul}
\usepackage{cite}
\makeatletter
\let\NAT@parse\undefined
\makeatother
\usepackage[breaklinks,colorlinks,linkcolor=black,citecolor=black,urlcolor=black]{hyperref}
\newtheorem{theorem}{Theorem}[section]
\newtheorem{lemma}[theorem]{Lemma}
\newtheorem{claim}[theorem]{Claim}
\newtheorem{corollary}[theorem]{Corollary}
\newtheorem{definition}[theorem]{Definition}

\acrodef{cfar}[CFAR]{constant false-alarm rate}
\acrodef{cpi}[CPI]{coherent processing interval}
\acrodef{cs}[CS]{Compressed sensing}
\acrodef{tbd}[TBD]{track-before-detect}
\acrodef{snr}[SNR]{signal-to-noise ratio}
\acrodef{ula}[ULA]{uniform linear array}
\acrodef{doa}[DoA]{direction of arrival}
\acrodef{pri}[PRI]{pulse repetition interval}
\acrodef{ctft}[CTFT]{continuous-time Fourier transform}
\acrodef{lasso}[LASSO]{least absolute shrinkage and selection operator}
\acrodef{awgn}[AWGN]{additive white Gaussian noise}
\acrodef{cwna}[CWNA]{continuous white noise acceleration}
\acrodef{ista}[ISTA]{iterative shrinkage-thresholding algorithm}
\acrodef{fista}[FISTA]{fast iterative shrinkage-thresholding algorithm}
\acrodef{dft}[DFT]{discrete Fourier transform}
\acrodef{gmm}[GMM]{Gaussian mixture model}
\acrodef{mf}[MF]{matched filtering}
\acrodef{mmv}[MMV]{multiple measurement vectors}
\acrodef{mimo}[MIMO]{multiple-input-multiple-output}
\acrodef{sbl}[SBL]{sparse Bayesian learning}
\acrodef{map}[MAP]{maximum a posteriori}
\acrodef{iht}[IHT]{iterative hard thresholding}
\acrodef{amp}[AMP]{approximate message passing}
\acrodef{rs}[RS]{replica symmetric}
\acrodef{tap}[TAP]{Thouless-Anderson-Palmer}
\acrodef{eccm}[ECCM]{electronic counter-countermeasures}
\acrodef{camp}[CAMP]{complex Approximate Message Passing}
\acrodef{glrt}[GLRT]{generalized likelihood ratio test}
\acrodef{mle}[MLE]{maximum likelihood estimation}
\acrodef{camp}[CAMP]{complex approximate message passing}
\acrodef{ecdf}[ECDF]{empirical cumulative distribution function}
\acrodef{crod}[CROD]{Complex Row-Orthogonal Debiased detector}
\acrodef{rod}[ROD]{Row-Orthogonal Debiased detector}
\acrodef{dct}[DCT]{discrete cosine transformation}
\acrodef{ks}[KS]{Kolmogorov-Smirnov}

\begin{document}
	\title{Compressed sensing radar detectors under the row-orthogonal design model: a statistical mechanics perspective}
	\author{Siqi~Na,
		Tianyao~Huang, 
		Yimin~Liu, 
		Takashi~Takahashi,\\ 
		Yoshiyuki~Kabashima, 
        and Xiqin Wang
    }
	\date{June 30, 2022}
	\maketitle
	
	\begin{abstract}
		\ac{cs} model of complex-valued data can represent the signal recovery process of a large amount types of radar systems, especially when the measurement matrix is row-orthogonal. 
		Based on debiased \ac{lasso}, detection problem under Gaussian random design model, i.e. the elements of measurement matrix are drawn from Gaussian distribution, is studied by literature. 
		However, we find that these approaches are not suitable for row-orthogonal measurement matrices. 
		In view of statistical mechanics approaches, we provide derivations of more accurate test statistics and thresholds (or p-values) under the row-orthogonal design model, and theoretically analyze the detection performance of the present detector. 
		Such detector can analytically provide the threshold according to given false alarm rate, which is not possible with the conventional \ac{cs} detector, and the detection performance is proved to be better than that of the traditional LASSO detector. 
		Comparing with other debiased \ac{lasso} based detectors, simulation results indicate that the proposed approach can achieve more accurate probability of false alarm when the measurement matrix is row-orthogonal, leading to better detection performance under Neyman-Pearson principle.  
	\end{abstract}
	
	\par\
	{\bf\emph{ Key words-\ Compressed sensing, radar detection, LASSO, row-orthogonal matrix, replica method, statistical mechanics}\rm}
	
	\tableofcontents
	
	\section{Introduction}
    \label{sec:intro}
    

    \ac{cs} model of complex-valued data assumes a scenario of recovering an $N$-dimensional vector ${\bm{x}}_0 \in {\mathbb{C}}^N$ from a $M$-dimensional vector ${\bm{y}} \in {\mathbb{C}}^M$, given by 
	\begin{equation}
	    \label{eq:csmodel}
	    {\bm{y}} = {\bm{A}}{\bm{x}}_0 + {\bm{\xi}},
	\end{equation}
	where ${\bm{A}} \in {\mathbb{C}}^{M\times N}$ is the measurement matrix (or the sensing matrix) and ${\bm{\xi}} \in {\mathbb{C}}^M$ is the complex \ac{awgn} with i.i.d. components $\xi_i \sim {\cal {CN}}\left( {0,\sigma^2} \right)$. 
	Basically, for the \ac{cs} model, $M$ is less than $N$, and we regard $\gamma = M / N$ as {\emph{compression rate}}. 
	The original signal ${\bm{x}}_0$ is a sparse vector containing only $k = 2 \rho N$ non-zero entries, where $2 \rho$ $(0 \le 2 \rho \le 1)$ is referred to as the {\emph{signal density}}. 
	
	In many practical applications, the signal processing can be modeled as above, while the sensing matrix has a specific structure. 
	Particularly, in this paper, we focus on radar applications, 
	in which ${\bm{y}}$, ${\bm{x}}_0$ and ${\bm{A}}$ refers to the sampled received signal, radar observation scene and observation (or steering) matrix, respectively.
	Generally, in the radar application scenarios, the number of non-zero entries in ${\bm{x}}_0$, which represents the intensity of scattering points, is small enough to satisfy $k \ll N$. 
	The indices of the non-zero entries indicate the ``position" of the scattering points (or the targets), such as range, azimuth, radial velocity, etc.
	
	The observation matrix $\bm A$ incorporates the geometry of the observation scene and the design of the transmitting waveform. 
	In the present paper, we are concerned with the case where $\bm A$ is row-orthogonal matrix.
	That is, ${\bm A}_{i \cdot} {\bm A}_{j \cdot}^H = 0$ for $i \ne j$ and ${\bm A}_{i \cdot} {\bm A}_{i \cdot}^H = 1$, in which ${\bm A}_{i \cdot}$ represents the $i$-th column of matrix $\bm A$. 
	A large amount types of radar transmitting waveform utilize a row-orthogonal steering matrix, and we list several below for illustration: 
	\begin{enumerate}[1)]
	    \item 
	    The partial observation problem of a pulse Doppler radar system, such as working in complex electromagnetic environments \cite{xiao2018distributed}, leads to a result of partial Fourier steering matrix, which is apparently row-orthogonal.
    	\item 
    	The steering matrix of frequency agile radar system is similarly row-orthogonal \cite{huang2018analysis, wang2021randomized}, which possesses a variety of merits such as good \ac{eccm} performance, low hardware system cost and convenience of spectrum sharing.
    	\item 
    	Sub-Nyquist radar systems \cite{cohen2018sub, na2018tendsur} realize the observation and compression on several dimensions, such as temporal domain, spatial domain and spectral domain.
    	The steering matrix of sub-Nyquist radar, which is proved to be the Kronecker product of several partial Fourier matrices (depends on the number of dimensions of the observation scene) in \cite{na2018tendsur}, satisfies the row-orthogonal property.
	\end{enumerate}
    
    For modern radar systems, it is crucial to detect targets element-wisely. 
    As previously mentioned, the indices of the non-zero entries in ${\bm{x}}_0$ represent the information of the targets, and judging whether each entry in ${\bm{x}}_0$ is non-zero can inform us about the existence of the targets and their location.
    We summarize such requirement as solving the following hypothesis testing problems: 
    \begin{equation}
        \label{eq:hypothesis_testing}
        \left\{ {\begin{array}{*{20}{l}}
        {{{\cal H}_{0, i}}:{x_{0,i}} = 0}, \\
        {{{\cal H}_{1, i}}:{x_{0,i}} \ne 0},
        \end{array}} \right.
    \end{equation}
    for $i = 1, 2, \ldots, N$ and designing thresholds for these tests. 
    Note that it is different from the task to detect whether there exists any target in the whole observed scene. The latter is generally more simple, which can be expressed as 
    \begin{equation}
        \label{eq:hypothesis_testing_any}
        \left\{ {\begin{array}{*{20}{l}}
        {{{\cal H}_{0}}:{\bm{y}} = {\bm{\xi}}}, \\
        {{{\cal H}_{1}}:{\bm{y}} = {\bm{A}}{\bm{x}}_0 + {\bm{\xi}}},
        \end{array}} \right.
    \end{equation}
    and can be easily solved by conventional approaches such as \ac{glrt} or Wald test. 
    While since the linear model \eqref{eq:csmodel} is underdetermined, the \ac{mle} of $x_{0, i}$ is unavailable.
    Consequently, one cannot directly apply conventional detectors to solve detection problem \eqref{eq:hypothesis_testing}.  
    
    Therefore, it is natural that we estimate ${\bm x}_0$ by \ac{cs} approaches and complete the tests on the basis of estimation, in which \ac{lasso} \cite{tibshirani1996regression} is a frequently used technique. 
	The complex-valued \ac{lasso} estimator $\hat {\bm{x}}^{\rm{LASSO}}$ is given by
	\begin{equation}
	    \hat {\bm{x}}^{\rm{LASSO}} = \mathop {\arg \min} \limits_{{\bm{x}}} \left\{ \frac{1}{2} {\left\| {\bm y} - {\bm A}{\bm x} \right\|_2^2} +  \lambda {\left\|{\bm x} \right\|_1} \right\},
	\end{equation}
	where 
	\begin{equation}
	    {\left\|{\bm x} \right\|_1} \buildrel \textstyle. \over = \sum\limits_{i = 1}^{N} {\sqrt{ \left( {\rm{Re}}\left( x_i \right) \right)^2 + \left( {\rm{Im}}\left( x_i \right) \right)^2}}.
	\end{equation}
	As a convex optimization problem, \ac{lasso} can be solved by standard techniques such as interior point and homotopy methods \cite{efron2004least, kim2007method}.
	There are also plenty of threshold class fast algorithms such as \ac{amp}, \ac{ista}, etc.
	Besides the computationally feasible feature, the nature of $\hat {\bm{x}}^{\rm{LASSO}}$ has been studied in a large literature, which mainly focus on: the prediction error $|| {\bm A} ( \hat {\bm{x}}^{\rm{LASSO}} - {\bm{x}}_0 ) ||_2^2/M$ \cite{greenshtein2004persistence}, the estimation error $|| \hat {\bm{x}}^{\rm{LASSO}} - {\bm{x}}_0 ||_q$ with $q \in \{1, 2\}$ \cite{bickel2009simultaneous, candes2007dantzig, raskutti2011minimax} and variable selection (or support estimation) of ${\bm{x}}_0$ \cite{meinshausen2006high, zhao2006model, wainwright2009sharp}, denoted by the support set $S_0 = \left\{ i \le N: x_{0, i} \ne 0 \right\}$ such that ${\mathbb{P}} (\hat{S} \ne S_0 )$ is bounded. 
	However, the above study is not enough to solve detection problems \eqref{eq:hypothesis_testing}: we still cannot get their p-values (or the threshold of given false alarm rate, which is extremely important in radar applications). 
	Naturally, one would wonder about the distribution of $\hat {x}^{\rm{LASSO}}_i$ in purpose of designing the detector.

    A certain linear transformation of the \ac{lasso} estimator, called {\emph{debiased \ac{lasso}}}, is given by
    \begin{equation}
        \label{eq:debiased_LASSO}
        {\hat {\bm x}}^{\rm {d}} = {\hat {\bm x}}^{\rm {LASSO}} + \frac{1}{\Lambda} {\bm A}^H ({\bm y} - {\bm A} {\hat {\bm x}}^{\rm {LASSO}}),
    \end{equation}
    where $\Lambda > 0$ is the {\emph{debiased coefficient}} computed from known variables and contains information about the structure of sensing matrix $\bm A$. 
    Asymptotic analysis of the \ac{lasso} solution based on \ac{amp} algorithm was developed in \cite{bayati2011dynamics} and \cite{bayati2011lasso}, which proves that the empirical distribution of the difference
    \begin{equation}
        \label{eq:w}
        {\bm{w}}  \buildrel \textstyle. \over = {\hat {\bm x}}^{\rm {d}} - {\bm x}_0,
    \end{equation}
    converges weakly to Gaussian distribution under Gaussian random design model (the entries of observation matrix $\bm A$ are i.i.d. drawn from Gaussian distribution). 
    This conclusion derives two studies for the design of detectors for solving the detection problem \eqref{eq:hypothesis_testing} based on debiased \ac{lasso}: \cite{javanmard2014hypothesis} for the real-valued \ac{cs} model and \cite{anitori2012design} for the complex-valued one. 
    While all the studies above restrict the measurement matrix to be Gaussian, work in \cite{takahashi2018statistical} based on statistical mechanics methods as well provides different ways for obtaining ${\hat {\bm x}}^{\rm {d}}$.
    The methodologies used allows the asymptotic analysis of \ac{lasso} solutions and the construction of debiased \ac{lasso} estimator for multiple real-valued observation matrix ensembles such as Gaussian, row-orthogonal, random \ac{dct}. 
    We aim to derive debiased \ac{lasso} in complex-valued form based on \cite{takahashi2018statistical}, especially under row-orthogonal design model, and analyze the detection performance of the resulting constructed detector. 
    
    
    The main contributions of this paper are twofold. 
    First, we present a general debiased \ac{lasso} detector framework and analyze its detection performance.
    We summarize the existing research on detection problems, including \cite{javanmard2014hypothesis} and \ac{camp} based \cite{anitori2012design}, and find that the test statistics used are the same: all debiased \ac{lasso}.
    Therefore, we construct a general detector framework based on debiased \ac{lasso}, and analyze its detection performance.
    We prove that the detection performance of debiased \ac{lasso} detector is better than that of traditional \ac{lasso} detector under the Neyman-Pearson principle, and its detection threshold can be analytically calculated by the given false alarm rate, thus the detection rate can be further quantified.
    In contrast, most \ac{cs} methods do not have closed-form solutions, and the distribution is not available.
    Consequently, the threshold cannot be given by the false alarm rate either.
    Therefore, we believe that the debiased lasso detector can completely replace the traditional \ac{cs} detector when solving the detection problem \eqref{eq:hypothesis_testing}.

    Second, we extend the results of  \cite{takahashi2018statistical} to the complex domain to enhance their applicability to engineering applications. 
    Some methods in statistical mechanics are used in reference \cite{takahashi2018statistical}, which sacrifice part of the mathematical rigor to obtain more attractive results: the debiased \ac{lasso} estimator for some non-Gaussian sensing matrices can be derived, and a more accurate estimation medium for the variance $\sigma_w^2$ of ${\bm{w}}$ (as defined in \eqref{eq:w}) can be provided.
    These results make it possible to implement debiased \ac{lasso} detector under row-orthogonal design model. 
    Therefore, for engineering applications, we extend these derivation processes to the complex domain.
    We find that the debiased coefficient under the row-orthogonal assumption is different from the results in \cite{javanmard2014hypothesis} and \cite{anitori2012design}, and the correctness of the results obtained by our method is verified by simulation experiments. 
    Numerical results also verify that our method can estimate the variance $\sigma_w^2$ of ${\bm{w}}$ more accurately than \cite{javanmard2014hypothesis} and \cite{anitori2012design}. 
    This leads to a more accurate threshold for the designed debiased \ac{lasso} detector and therefore to a better reaching of a given false alarm rate. 
    Such conclusion is also verified by simulation results.

	The organization of the present paper is as follows. 
	In Section \ref{sec:method}, we provide our results on the design of the detector for compressed sensing radar system and the analysis on its detection performance. 
	The derivation of the test statistic and the threshold of the presented detector is elaborated in Section \ref{sec:analysis}. 
	Section \ref{sec:result} furnishes some numerical validation of the previous asymptotic analysis. 
	We conclude the paper in Section \ref{sec:conclusion}. 
	
	Throughout the paper, we use $a$, $\bm a$ and $\bm A$ as a number, a vector and a matrix, respectively. 
	For a set $S$, $\# S$ denotes its cardinality. 
	Denote by ${\rm{Re}}(\cdot)$ and ${\rm{Im}}(\cdot)$ for the real and imaginary parts of a complex-valued component, respectively.
	Function $\delta(\cdot)$ is the Dirac's delta function and $\Theta (x)$ is Heaviside's step function.
	The operators $(\cdot)^T$, $(\cdot)^*$ and $(\cdot)^H$ represent the transpose, conjugate and conjugate transpose of a component, respectively. 
	Denote by $\varphi(x) = {\rm{e}}^{-x^2/2}/\sqrt{2\pi}$ the Gaussian density and $\Phi(x) = \int_{-\infty}^x \varphi(u) {\rm{d}}u$ the Gaussian distribution.
	
    \section{Design and analysis of the debiased LASSO detector}
    \label{sec:method}
    
    In this section, we first introduce the debiased \ac{lasso} detector. 
    The advantage of the debiased \ac{lasso} detector is elaborated mainly by comparing with the \ac{lasso} detector in Section \ref{subsec:debiased_LASSO_detector}.
    While both the proposed detector and the ones in  \cite{javanmard2014hypothesis, anitori2012design, takahashi2018statistical} are debiased \ac{lasso} detectors, the difference is explained in Section \ref{subsec:differences}. 
    Particularly, this paper inherits some derivation from \cite{takahashi2018statistical}, and we will briefly describe the contribution of this work over \cite{takahashi2018statistical} in Section \ref{subsec:differences}.
    	
	\subsection{Debiased LASSO detector}
	\label{subsec:debiased_LASSO_detector}
	    
    	\begin{figure}
    	    \centering
    	    \includegraphics[width=14cm]{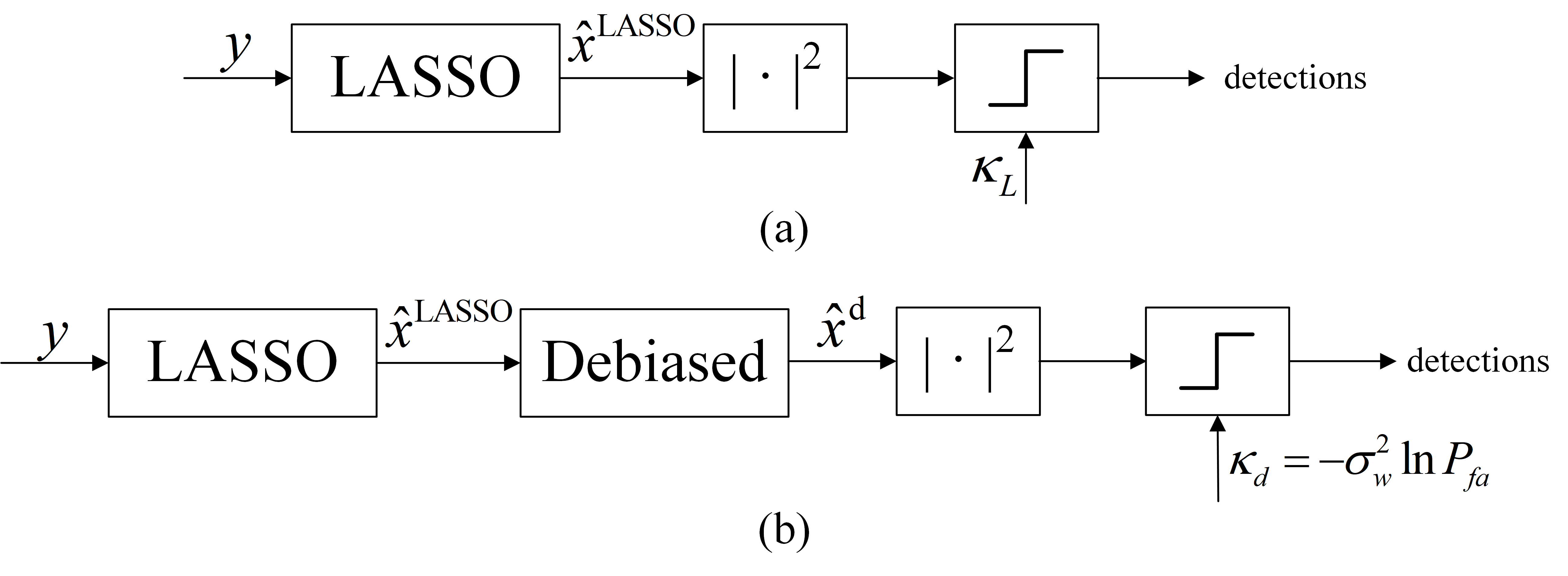}
    	    \caption{Frameworks of two detectors for detection problem \eqref{eq:hypothesis_testing}: (a) \ac{lasso} detector; (b) Debiased \ac{lasso} detector. }
    	    \label{fig:Debiased_LASSO_detector_framework}
    	\end{figure}
    	
    	
    	With regard to detection problem \eqref{eq:hypothesis_testing}, we here define the following performance metrics.
	    Denote by
	    \begin{equation}
	        {\varphi_i} = \left\{ {\begin{array}{*{20}{c}}
            {1,}&{{\text{if detector reject }}{H_{0,i}};}\\
            {0,}&{{\text{otherwise.}}}
            \end{array}} \right.
	    \end{equation}
	    Define the false alarm probability $P_{fa}$ as
	    \begin{equation}
	        P_{fa} = \mathop {\lim }\limits_{N \to \infty } \frac{1}{N-k} \sum\limits_{i \in S^c} { \varphi_{i} },
	    \end{equation}
	    and the detection probability $P_{d}$ as
	    \begin{equation}
	        P_{d} = \mathop {\lim }\limits_{N \to \infty } \frac{1}{k} \sum\limits_{i \in S} { \varphi_{i} },
	    \end{equation}
	    where $S = \left\{ i: x_{0, i} \ne 0 \right\}$ is the support set of ${\bm x}_0$ with $\# S = k$ and $S^c = \{1, \ldots, N\} \backslash S$.

    	The \ac{lasso} detector, which is a natural idea, possesses the structure shown in Fig. \ref{fig:Debiased_LASSO_detector_framework} (a), 
    	in which the test statistic is provided by \ac{lasso} estimator. 
    	Indeed, other traditional \ac{cs} radar detectors have the similar form with the \ac{lasso} detector, which contain the compressed sensing module and a test of judging whether the amplitude is greater than a threshold. 
    	
    	On the contrary, the debiased \ac{lasso} detector for compressed sensing radar has the structure shown in Fig. \ref{fig:Debiased_LASSO_detector_framework} (b). 
    	The test statistic is given by debiased estimator ${\hat {\bm x}}^{\rm {d}}$ and the threshold $\kappa_d$ can fix the probability of false alarm. 
    	In the present paper, we propose the debiased \ac{lasso} detector for complex-valued row-orthogonal observation matrix called \ac{crod}, whose procedure is shown in Algorithm \ref{alg:CROD}. 
    	The detailed derivation of the involved parameters will be presented later in Section \ref{sec:analysis}.

	    We declare here that the benefits of the debiased LASSO detector are mainly twofold. 
	    First, the relationship between the threshold $\kappa_d$ and the false alarm rate $P_{fa}$ can be computed analytically, as will be detailed later in Section \ref{subsubsec:analytical}. 
	    This is usually not possible for traditional \ac{cs} detectors, because the distribution of the solution obtained by \ac{cs} methods is unavailable. 
	    Second, its detection performance is better than that of the \ac{lasso} detector, which will be proved in Section \ref{subsubsec:detection_performance}. 
    	
    	\begin{algorithm}
            \caption{\ac{crod}}
            \label{alg:CROD}
            \begin{algorithmic}[1]
                \Require $\bm y$, $\bm A$, regularization parameter $\lambda$, probability of false alarm $P_{fa}$, variance of noise $\sigma^2$
                \Ensure debiased \ac{lasso} estimator ${\hat {\bm x}}^{\rm {d}}$, threshold $\kappa_d$
                \State Let 
                    \begin{equation}
                	    \hat {\bm{x}}^{\rm{LASSO}} = \mathop {\arg \min} \limits_{{\bm{x}}} \left\{ \frac{1}{2} {\left\| {\bm y} - {\bm A}{\bm x} \right\|_2^2} +  \lambda {\left\|{\bm x} \right\|_1} \right\}.
                	    \nonumber
                	\end{equation}
            	\State The debiased \ac{lasso} estimator is obtained from
                    \begin{equation}
                	    \label{eq:debiasedLASSO}
            	        {\hat {\bm x}}^{\rm {d}} = {\hat {\bm x}}^{\rm {LASSO}} + \frac{1}{\Lambda_{\rm{CROM}}} {\bm A}^H ({\bm y} - {\bm A} {\hat {\bm x}}^{\rm {LASSO}}),
            	    \end{equation}
            	    with $\Lambda_{\rm{ROM}}$ and $\rho_{\rm{CA}}$ given by
                    \begin{eqnarray}
                        &&\Lambda_{\rm{CROM}} = \frac{\gamma - \rho_{\rm{CA}}}{1 - \rho_{\rm{CA}}}, \nonumber \\
                        \label{eq:alg_rho_CA}
            	        &&\rho_{\rm{CA}} = \frac{1}{2N} {\sum \limits_{i = 1}^N \left[ \left( 2 - \frac{\lambda}{\Lambda_{\rm{CROM}} \left| {\hat x_i^{\rm{LASSO}}} \right| + \lambda } \right) \cdot  \Theta \left( \left| {\hat x_i^{\rm{LASSO}}} \right| \right) \right] }. 
                    \end{eqnarray}
                \State The threshold $\kappa_d$ is given by
                    \begin{equation}
                        \kappa_d = -\sigma_w^2 \ln{P_{fa}}, \nonumber
                    \end{equation}
                    where
        	        \begin{eqnarray}
        	            \label{eq:alg_chi}
                        &&\chi = \frac{\rho_{\rm{CA}} (1 - \rho_{\rm{CA}})}{\gamma - \rho_{\rm{CA}}},  \\
                        \label{eq:alg_G'}
                        &&G'(-\chi; {\bm J}) = \frac{1 + \chi - \sqrt{(\chi+1)^2 - 4 \gamma \chi}}{2\chi},  \\
                        \label{eq:alg_G''}
                        &&G''(-\chi; {\bm J}) = \frac{2 \gamma \chi - \chi - 1 + \sqrt{(\chi+1)^2 - 4 \gamma \chi}}{2 \chi^2 \sqrt{(\chi+1)^2 - 4 \gamma \chi}},  \\
                        \label{eq:alg_RSS}
                        &&\overline{\rm{RSS}} = \frac{1}{M}  \left\| \bm{y} - \bm{A} {\hat {\bm x}}^{\rm {LASSO}} \right\|_2^2,  \\
                        \label{eq:alg_hatchi}
                        &&\hat \chi = \frac{\gamma G''(-\chi; \bm{J})}{2 G'(-\chi; \bm{J}) - 2 G''(-\chi; \bm{J}) \chi} \overline{\rm{RSS}} + \frac{-G''(-\chi; \bm{J}) \gamma + \left( G'(-\chi; \bm{J}) \right)^2}{2 G'(-\chi; \bm{J}) - 2 G''(-\chi; \bm{J}) \chi} \sigma^2,  \\
                        \label{eq:alg_sigma}
                        &&\sigma_w^2 = \frac{2 \hat \chi}{\Lambda_{\rm{CROM}}^2}. 
                    \end{eqnarray}
                
            \end{algorithmic}
        \end{algorithm}
    	
    	\subsubsection{Analytical expression of false alarm probability}
    	\label{subsubsec:analytical}
    	
    	The most significant advantage of the debiased \ac{lasso} detector over the conventional \ac{cs} detector is that the analytical relationship between the threshold $\kappa_d$ and the false alarm rate $P_{fa}$ can be obtained. 
    	Controlling false alarm rate is particularly important in many radar applications due to resource allocation and other reasons. 
	    Asymptotic analysis shows that as $N \to \infty$, if $\Lambda$ is suitably chosen, ${\hat x}^{\rm {d}}_i$, the $i$-th entry of ${\hat {\bm x}}^{\rm {d}}$, approximately follows Gaussian distribution with mean $x_{0, i}$, where $x_{0, i}$ is the $i$-th entry of ${\bm x}_0$. 
	    For the convenience of readers, we here restate the definition and conclusion in \cite{bayati2011dynamics} (, which is similar to that in \cite{bayati2011lasso},) to describe this result more precisely. 
	    
	    \begin{definition}[\cite{bayati2011dynamics}]
	        For a given $(\gamma, 2\rho) \in [0, 1]^2$, a sequence of instances $\{ {\bm{x}}_0(N), {\bm{\xi}}(N), {\bm{A}}(N) \}_{N \in \mathbb{N}}$ indexed by $N$ is said to be a converging sequence of Gaussian design model if the empirical distribution of the entries ${\bm{x}}_0(N) \in \mathbb{R}^N$ converges weakly to a probability measure $p_X$ with bounded second moment, the empirical distribution of the entries ${\bm{\xi}}(N) \in \mathbb{R}^M$ $(M = \gamma N)$ converges weakly to a probability measure $p_{\xi}$ with bounded second moment and the elements of ${\bm{A}}(N) \in \mathbb{R}^{M \times N}$ are i.i.d. drawn from a Gaussian distribution.
	    \end{definition}
	    
	    \begin{lemma}[\cite{bayati2011dynamics}]
	        \label{lemma:converge_Gaussian}
	        Let $\{ {\bm{x}}_0(N), {\bm{\xi}}(N), {\bm{A}}(N) \}_{N \in \mathbb{N}}$ be a converging sequence of Gaussian design model. 
	        The empirical law of ${\bm{w}}(N) = {\hat {\bm x}}^{\rm {d}}(N) - {\bm{x}}_0(N)$ converges to a zero-mean Gaussian distribution almost surely as $N \to \infty$ for a specific $\Lambda$. 
	    \end{lemma}
	    
	    In the present paper, we further prove that such $\Lambda$ is unique. 
	    
	    \begin{theorem}
            \label{theorem:lambda_unique}
            Let $\{ {\bm{x}}_0(N), {\bm{\xi}}(N), {\bm{A}}(N) \}_{N \in \mathbb{N}}$ be a converging sequence of Gaussian design model.
            Let $\Lambda_1 \in {\mathbb{R}}$ and ${\hat {\bm x}}^{\rm {d}}_1 = {\hat {\bm x}}^{\rm {LASSO}} + \frac{1}{\Lambda_1} {\bm A}^H ({\bm y} - {\bm A} {\hat {\bm x}}^{\rm {LASSO}})$, such that the empirical law of ${\bm{w}}_1(N) = {\hat {\bm x}}^{\rm {d}}_1(N) - {\bm{x}}_0(N)$ converges to a zero-mean Gaussian distribution almost surely as $N \to \infty$.
            Then for all $\Lambda_2 \ne \Lambda_1$ and ${\hat {\bm x}}^{\rm {d}}_2 = {\hat {\bm x}}^{\rm {LASSO}} + \frac{1}{\Lambda_2} {\bm A}^H ({\bm y} - {\bm A} {\hat {\bm x}}^{\rm {LASSO}})$, the empirical law of ${\bm{w}}_2(N) = {\hat {\bm x}}^{\rm {d}}_2(N) - {\bm{x}}_0(N)$ does not converge to a Gaussian distribution as $N \to \infty$.
            \begin{proof}
                The proof is given in Appendix \ref{appendix:unique}.
            \end{proof}
        \end{theorem}
	    
	    In addition, we claim that such uniqueness is not limited to Gaussian sensing matrix, but other matrix manifolds also have the same conclusion such as row-orthogonal matrices. 
	    The following definition of a converging sequence of row-orthogonal design model is required. 
        \begin{definition}
	        For a given $(\gamma, 2\rho) \in [0, 1]^2$, a sequence of instances $\{ {\bm{x}}_0(N), {\bm{\xi}}(N), {\bm{A}}(N) \}_{N \in \mathbb{N}}$ indexed by $N$ is said to be a converging sequence of row-orthogonal design model if the empirical distribution of the entries ${\bm{x}}_0(N) \in \mathbb{R}^N$ converges weakly to a probability measure $p_X$ with bounded second moment, the empirical distribution of the entries ${\bm{\xi}}(N) \in \mathbb{R}^M$ $(M = \gamma N)$ converges weakly to a probability measure $p_{\xi}$ with bounded second moment and ${\bm{A}}(N) \in \mathbb{R}^{M \times N}$ is randomly drawn from all the row-orthogonal matrices with a size of $M \times N$.
	    \end{definition}
        
        \begin{corollary}
            \label{corollary:lambda_unique}
            Let $\{ {\bm{x}}_0(N), {\bm{\xi}}(N), {\bm{A}}(N) \}_{N \in \mathbb{N}}$ be a converging sequence of row-orthogonal design model.
            Let $\Lambda_1 > 0$ and ${\hat {\bm x}}^{\rm {d}}_1 = {\hat {\bm x}}^{\rm {LASSO}} + \frac{1}{\Lambda_1} {\bm A}^H ({\bm y} - {\bm A} {\hat {\bm x}}^{\rm {LASSO}})$, such that the empirical law of ${\bm{w}}_1(N) = {\hat {\bm x}}^{\rm {d}}_1(N) - {\bm{x}}_0(N)$ converges to a zero-mean Gaussian distribution almost surely as $N \to \infty$.
            Then for all $\Lambda_2 \ne \Lambda_1$ and ${\hat {\bm x}}^{\rm {d}}_2 = {\hat {\bm x}}^{\rm {LASSO}} + \frac{1}{\Lambda_2} {\bm A}^H ({\bm y} - {\bm A} {\hat {\bm x}}^{\rm {LASSO}})$, the empirical law of ${\bm{w}}_2(N) = {\hat {\bm x}}^{\rm {d}}_2(N) - {\bm{x}}_0(N)$ does not converge to a Gaussian distribution as $N \to \infty$.
        \end{corollary}
        
        The proof of the corollary is the same as Theorem \ref{theorem:lambda_unique}. 
	    
	    Denote the sample variance of ${\bm{w}}(N)$ by $\sigma_w^2$, it is natural to get the following analytical relationship between the probability of false alarm and the threshold $\kappa_d$ of the detector.
	    
	    \begin{theorem}
	        \label{theorem:pfa_kappa}
	        The false alarm probability $P_{fa}$ of the debiased \ac{lasso} detector satisfies: 
	        \begin{equation}
	            \label{eq:pfa_kappa}
	            \kappa_d = - \sigma_w^2 \ln P_{fa},
	        \end{equation}
	        where the test is
	        \begin{equation}
    	        {\varphi_{i}} = \left\{ {\begin{array}{*{20}{l}}
                {1,}&{\left| \hat {x}^{\rm{d}}_i \right|^2 > \kappa_d ;}\\
                {0,}&{{\text{otherwise.}}}
                \end{array}} \right.
    	    \end{equation}
            \begin{proof}
                For $i \in S^c$, which means that $x_{0, i} = 0$, the empirical law of $\{ \hat x^{\rm{d}}_i \}$ converges to ${\cal{CN}} (0, \sigma_w^2)$ as $N \to \infty$.
                Therefore, the empirical distribution of $ \{ \left| \hat {x}^{\rm{d}}_i \right|^2 \}$ converges to exponential distribution of rate parameter $1/\sigma_w^2$, leading to: 
                \begin{equation}
                    P_{fa} = \mathop {\lim }\limits_{N \to \infty } \frac{1}{N-k} \sum\limits_{i \in S^c} { \varphi_{i} } = \exp{\left( - \frac{\kappa_d}{\sigma_w^2} \right)},
                \end{equation}
                which proves \eqref{eq:pfa_kappa}.
            \end{proof} 
	    \end{theorem}
	    
	    From Theorem \ref{theorem:pfa_kappa}, we conclude that the threshold $\kappa_d$ of the debiased \ac{lasso} detector can be analytically calculated by the false alarm probability.
	    Such mission cannot be achieved by traditional \ac{cs} detector for the reason that almost all of the solutions obtained by \ac{cs} methods does not have a closed form, neither the distribution of the solutions.
	    In the present paper, we will provide an asymptotic analysis from a statistical mechanics perspective and the process of calculating the coefficients $\Lambda$ with variance $\sigma_w^2$ in Section \ref{sec:analysis}, which can be adapted to the row-orthogonal matrix design of $\bm A$ mentioned for radar applications. 
	    
	    \subsubsection{Better detection performance}
    	\label{subsubsec:detection_performance}
	    
	    At first glance, the debiased \ac{lasso} estimator destroys the sparsity brought by the \ac{lasso} results, but we will next prove theoretically that treating $\left| {\hat x}^{\rm {d}}_i \right|$ as a test statistic compared to $\left| {\hat x}^{\rm {LASSO}}_i \right|$ will improve the detection performance of the detector. 
	    
	    \begin{theorem}
	    
	        \label{theorem:detection_performance}
	        
	        For the same false alarm probability $P_{fa}$, the detection probability of debiased \ac{lasso} detector $P_{d, 1}$ is not less than that of \ac{lasso} detector $P_{d, 2}$.
	        
	        \begin{proof}
	            See Appendix \ref{appendix:performance} for the proof.
	        \end{proof}

	    \end{theorem}

	    Theorem \ref{theorem:detection_performance} suggests that applying such non-sparse solution in the detector instead lead to better detection performance.

    \subsection{Comparison with existing debiased \ac{lasso} detectors}
    \label{subsec:differences}
    
        \begin{table} 
            \caption{Comparison of several detectors.}
            \begin{center}  
                \begin{tabular}{cccccc} 
                    \hline 
                    \textbf{approach} & $\Lambda$ & $\sigma_w^2$ & \textbf{Gaussian} & \textbf{row-orthogonal} & \textbf{complex}  \\
                    \hline 
                    SDL-test \cite{javanmard2014hypothesis} & \eqref{eq:Lambda_G} & \eqref{eq:variance_SDL} & \checkmark & $\times$ & $\times$ \\
                    CAMP \cite{anitori2012design} & \eqref{eq:Lambda_CG} & \eqref{eq:variance_CAMP} & \checkmark & $\times$ & \checkmark \\
                    ROD \cite{takahashi2018statistical} & \eqref{eq:Lambda_ROM} & \eqref{eq:alg_chi} - \eqref{eq:alg_sigma} & \checkmark & \checkmark & $\times$ \\
                    CROD & \eqref{eq:Lambda_CROM} & \eqref{eq:alg_chi} - \eqref{eq:alg_sigma} & \checkmark & \checkmark & \checkmark \\
                    \hline 
                \end{tabular}  
            \end{center}
            \label{tab:detectors_comparison} 
        \end{table}
        
        In this subsection, we compare the proposed detector \ac{crod} with SDL-test \cite{javanmard2014hypothesis}, \ac{camp} \cite{anitori2012design}, and the \ac{rod} constructed by the conclusions obtained from \cite{takahashi2018statistical}. 
        Due to the frameworks of all the debiased \ac{lasso} detectors are the same, we mainly list the differences of the debiased coefficient $\Lambda$ and the approaches to estimate the variance $\sigma_w^2$. 
        
        \begin{enumerate}
        
            \item 
            SDL-test \cite{javanmard2014hypothesis} suggests that
            \begin{equation}
                \label{eq:Lambda_G}
                \Lambda_{\rm{G}} = \gamma - \rho_a, 
            \end{equation}
            where $\rho_a = \# \{ i | {\hat x_i^{\rm{LASSO}}} \ne 0 \}/N$ denotes the {\emph{active component density}} and
            \begin{equation}
                \label{eq:variance_SDL}
                \hat \sigma_w = \frac{\sqrt{\gamma}}{\Phi^{-1} (0.75) (\gamma - \rho_a)} {\rm{median}} \left( \left| {\bm y} - {\bm A} {\hat {\bm x}}^{\rm {LASSO}} \right| \right), 
            \end{equation}
            for real-valued Gaussian random sensing matrix.
            
            \item
            \ac{camp} \cite{anitori2012design} provides 
            \begin{equation}
                \label{eq:Lambda_CG}
                \Lambda_{\rm{CG}} = \gamma - \rho_{\rm{CA}}, 
            \end{equation}
            where $\rho_{\rm{CA}}$ given by \eqref{eq:alg_rho_CA} denotes the {\emph{active component density}} and
            \begin{equation}
                \label{eq:variance_CAMP}
                \hat \sigma_w = \frac{1}{\sqrt{\ln 2}} {\rm{median}} \left( \left| {\hat {\bm x}}^{\rm {d}} \right| \right). 
            \end{equation}
            for complex-valued Gaussian random sensing matrix. 
            
            \item
            \ac{rod} from \cite{takahashi2018statistical} claims that 
            \begin{equation}
                \label{eq:Lambda_ROM}
                \Lambda_{\rm{ROM}} = \frac{\gamma - \rho_a}{1 - \rho_a}, 
            \end{equation}
            and $\sigma_w^2$ is given by exchanging all the $\rho_{\rm{CA}}$ in \eqref{eq:alg_chi} - \eqref{eq:alg_sigma} to $\rho_a$ for real-valued row-orthogonal sensing matrix. 
            
            \item
            The proposed \ac{crod} claims that 
            \begin{equation}
                \label{eq:Lambda_CROM}
                \Lambda_{\rm{CROM}} = \frac{\gamma - \rho_{\rm{CA}}}{1 - \rho_{\rm{CA}}}, 
            \end{equation}
            and $\sigma_w^2$ is given by \eqref{eq:alg_chi} - \eqref{eq:alg_sigma} for complex-valued row-orthogonal sensing matrix. 
            
        \end{enumerate}
        
        Moreover, the methodologies in \cite{takahashi2018statistical} can obtain the same result as \eqref{eq:Lambda_G} under Gaussian random matrix design. 
        Our work also gives the same debiased coefficient as \eqref{eq:Lambda_CG} for complex-valued Gaussian sensing matrix in Corollary \ref{corollary:debiased_Gaussian}. 
        Based on the derivation in Section \ref{sec:analysis}, we believe that $\Lambda$ is related to the asymptotic eigenvalue distribution $\rho_{\bm J}(s)$ of ${\bm J} = {\bm A}^T {\bm A}$ (or ${\bm J} = {\bm A}^H {\bm A}$), which is referred to Claim \ref{claim:debiased}. 
        For a Gaussian matrix, $\rho_{\bm J}(s)$ can be calculated by \eqref{eq:rho_J_Gaussian}, which yields \eqref{eq:Lambda_CG}. 
        For other sensing matrices, the proposed method is also applicable as $\rho_{\bm J}(s)$ is obtainable. 
        We summarize the results of the comparison in Table \ref{tab:detectors_comparison}.
        
        Recall that we have proved the uniqueness of the debiased coefficient $\Lambda$. 
        When the steering matrix $\bm A$ is row-orthogonal, the correctness of the debiased coefficient $\Lambda_{\rm{CROM}}$ given in the present paper will be verified by numerical result in Section \ref{sec:result}.
        Besides the construction of the debiased \ac{lasso} estimator, the approach to estimate $\sigma_w^2$ in the present paper is different from \ac{camp} and SDL-test. 
        There is also numerical result of the accuracy of the three estimation mediums presented in Section \ref{sec:result}.
        
        
        The derivation, including the construction of the debiased \ac{lasso} estimator and the estimation of the variance, in Section \ref{sec:analysis} is primarily based on \cite{takahashi2018statistical}. 
        However, since the signal model in radar systems are composed of complex vectors or matrices, our next derivations are in complex form, which differs from both the process and the results of \cite{takahashi2018statistical}. 
        We will verify in our simulation experiments that the direct use of the real-valued version results is incorrect. 
        This is because the complex \ac{lasso} differs from the real one with regard to the computation of the $\ell_1$-norm of $\bm x$ in the regularization term. 
    
    \section{Derivation of the debiased LASSO estimator and the estimation of its variance}
    \label{sec:analysis}
    
    In Algorithm 1, we simply present the results of the debiased 
    \ac{lasso} estimator and estimation of its variance $\sigma_w^2$. 
    In this section, we will provide the detailed derivation of these results, based on some statistical mechanics approaches. 
    Led by \cite{tanaka2002statistical}, more and more work apply statistical mechanics approaches for theoretical analysis in information theory and communications theory.
    Although a mathematically rigorous justification of replica method is still in-progress, it has been proved extensively successful in very difficult problems and applied to derive a number of captivating results, see \cite{guo2005randomly, tanaka2005approximate, campo2011large, wu2012optimal}.
    In addition, we refer to the results in this Section as claims. 
    
    Particularly, the asymptotic analysis of the distribution of LASSO solutions will first be introduced, which tells us that a central limit obeying Gaussian distribution with a mean of ${\bm x}_0$ exists.
    Then, we aim to get the close-form expression of the central limit, which will be regarded as the debiased LASSO estimator.
    At the end of this section, the procedure for estimating the variance of the debiased estimator will be presented.
    
    Throughout the analysis, we will generally examine the typical reconstruction performance in the limit $N, M \to \infty$, but keeping the compression rate $\gamma$ constant. 
    The assumption of large system limit $N \to \infty$ will be omitted in the following part to avoid wording.
    We list our main results in Section \ref{subsec:main_results}. 
    The proof details are referred to Section \ref{subsec:distribution}, \ref{subsec:debiased} and \ref{subsec:variance}.
    
    \subsection{Main results}
	\label{subsec:main_results}
	    
	    Our first result provides the asymptotic distribution of the \ac{lasso} solution $\hat {\bm{x}}^{\rm{LASSO}}$. 
	    \begin{claim}
	        \label{claim:distribution}
	        When the sensing matrix $\bm A$ is randomly drawn from row-orthogonal matrices, the asymptotic distribution of the LASSO solution $\hat {\bm{x}}^{\rm{LASSO}}$ can be inferred as follow:
    	    \begin{equation}
    	        \hat x_i^{\rm{LASSO}} =  {\rm{ST}}_{\lambda, \hat Q} \left( h_i \right) \buildrel \textstyle. \over = \frac{h_i}{\left| h_i \right|} \cdot \frac{\left| h_i \right| - \lambda}{\hat Q} \Theta \left( \left| h_i \right| - \lambda \right), 
    	    \end{equation}
    	    where 
    	    \begin{equation}
    	        \label{eq:h_i}
    	        h_i = \hat m{x_{0,i}} + \sqrt {2 \hat \chi } {z_i},
    	    \end{equation}
    	    in which $z_i \sim {\cal{CN}} (0, 1)$ are i.i.d. standard complex variables and $\hat m$, $\hat Q$, $\hat \chi$ are positive real numbers. 
	    \end{claim}
	    Such conclusion reminds us that the debiased \ac{lasso} estimator can be easily obtained if $h_i$ is available.
	    We next introduce the construction of debiased \ac{lasso} ${\hat {\bm x}}^{\rm {d}}$ for certain matrix ensembles through deriving $h_i$.
	    \begin{claim}
	        \label{claim:debiased}
	        Suppose the Gram matrix $ {\bm J} = {\bm A}^H {\bm A}$ has deterministic asymptotic eigenvalue distribution $\rho_{\bm J}(s)$, the debiased \ac{lasso} estimator ${\hat {\bm x}}^{\rm {d}}$ is given by
    	    \begin{equation}
    	        \label{eq:debiased}
    	        {\hat {\bm x}}^{\rm {d}} = {\hat {\bm x}}^{\rm {LASSO}} + \frac{1}{\Lambda} {\bm A}^H ({\bm y} - {\bm A} {\hat {\bm x}}^{\rm {LASSO}}). 
    	    \end{equation}
    	    The debiased coefficient is 
    	    \begin{equation}
    	        \label{eq:Lambda}
    	        \Lambda = \frac{t \cdot \rho_{\rm{CA}}}{\rho_{\rm{CA}} - 1},
    	    \end{equation}
    	    in which $t$ is the solution of 
    	    \begin{equation}
    	        \label{eq:eq_t}
    	        \int \frac{\rho_{\bm J}(s)}{t-s} {\rm{d}}s = \frac{1 - \rho_{\rm{CA}}}{t}, 
    	    \end{equation}
    	    and $\rho_{\rm{CA}}$ is complex active component density of the \ac{lasso} solution defined in \eqref{eq:alg_rho_CA}. 
	    \end{claim}
	    
	    Setting the sensing matrix ensemble to row-orthogonal, then the asymptotic eigenvalue distribution of $\bm J$ is given by
	    \begin{equation}
	        \label{eq:rhoJs}
	        \rho_{\bm J}(s) = (1 - \gamma) \delta(s) + \gamma \delta(s - 1).
	    \end{equation}
	    Therefore, one can obtain the following corollary. 
	    \begin{corollary}
	        \label{corollary:debiased_PF}
	        When the complex-valued sensing matrix $\bm A$ is row-orthogonal, the debiased \ac{lasso} estimator ${\hat {\bm x}}^{\rm {d}}$ is given by
    	    \begin{equation}
    	        \label{eq:debiased_PF}
    	        {\hat {\bm x}}^{\rm {d}} = {\hat {\bm x}}^{\rm {LASSO}} + \frac{1}{\Lambda_{\rm{CROM}}} {\bm A}^H ({\bm y} - {\bm A} {\hat {\bm x}}^{\rm {LASSO}}),
    	    \end{equation}
    	    where
    	    \begin{equation}
    	        \Lambda_{\rm{CROM}} = \frac{\gamma - \rho_{\rm{CA}}}{1 - \rho_{\rm{CA}}},
    	    \end{equation}
    	    and $\rho_{\rm{CA}}$ is complex active component density of the \ac{lasso} solution defined in \eqref{eq:alg_rho_CA}.
	    \end{corollary}
	    
	    According to \cite{marvcenko1967distribution}, when the entries of $\bm A$ are all i.i.d. Gaussian ensembles with mean $0$ and variance $1/N$, the asymptotic eigenvalue distribution is given by
	    \begin{eqnarray}
	        \label{eq:rho_J_Gaussian}
	        &&\rho_{\bm J}(s) = (1 - \gamma) \delta(s) + \frac{1}{2\pi} \frac{\sqrt{(\lambda_+ - s)(s - \lambda_-)}}{s} {\mathbb{I}}_{[\lambda_-, \lambda_+]}(s), \\
	        &&\lambda_{\pm} = (1 \pm \sqrt{\gamma})^2, \\
	        &&{\mathbb{I}}_{S}(x) = \left\{ {\begin{array}{*{20}{c}}
            1&{{\text{if }} x \in S}\\
            0&{{\text{otherwise}}}
            \end{array}} \right. .
	    \end{eqnarray}
	    Then comes the following corollary. 
        \begin{corollary}
	        \label{corollary:debiased_Gaussian}
	        When the sensing matrix $\bm A$ is of random i.i.d. complex Gaussian ensemble in which all entries of $\bm A$ are i.i.d. complex Gaussian variables with mean $0$ and variance $1/N$, the debiased LASSO estimator ${\hat {\bm x}}^{\rm {d}}$ is given by
    	    \begin{equation}
    	        \label{eq:debiasedGaussian}
    	        {\hat {\bm x}}^{\rm {d}} = {\hat {\bm x}}^{\rm {LASSO}} + \frac{1}{\Lambda_{\rm{CG}}} {\bm A}^H ({\bm y} - {\bm A} {\hat {\bm x}}^{\rm {LASSO}}),
    	    \end{equation}
    	    where $\Lambda_{\rm{CG}} = \gamma - \rho_{\rm{CA}}$.
	    \end{corollary}
	    In addition, we find that the debiased coefficient $\Lambda_{\rm{CG}}$ is the same as the result in \cite{anitori2012design} in the case of complex Gaussian matrix design. 
	    
	    The third result presents a method for estimating the variance $\sigma_w^2$. 
	    \begin{claim}
            \label{claim:variance}
            When the sensing matrix $\bm A$ is randomly drawn from row-orthogonal matrices, the sample variance $\sigma_w^2$ of ${\bm w} = {\hat {\bm x}}^{\rm {d}} - {\bm x}_0$ converges to $2 \hat \chi / \hat Q^2$, where 
            \begin{equation}
                \hat Q = G'(-\chi; {\bm J}),
            \end{equation}
            and
            \begin{equation}
                \label{eq:eqhatchi}
                \hat \chi = \frac{\gamma G''(-\chi; \bm{J})}{2 G'(-\chi; \bm{J}) - 2 G''(-\chi; \bm{J}) \chi} \overline{\rm{RSS}} + \frac{-G''(-\chi; \bm{J}) \gamma + \left( G'(-\chi; \bm{J}) \right)^2}{2 G'(-\chi; \bm{J}) - 2 G''(-\chi; \bm{J}) \chi} \sigma^2.
            \end{equation}
            Here,  
            \begin{eqnarray}
                \label{eq:Gprimechi}
                &&G'(-\chi; {\bm J}) = t(-\chi) + \frac{1}{\chi}, \\
                \label{eq:Gwprimechi}
                &&G''(-\chi; {\bm J}) = t'(-\chi) + \frac{1}{\chi^2}, \\
                &&\overline{\rm{RSS}} = \mathop {\lim }\limits_{N \to \infty } \frac{1}{N \gamma}  \left\| \bm{y} - \bm{A} {\hat {\bm x}}^{\rm {LASSO}} \right\|_2^2.
            \end{eqnarray}
            in which
            \begin{eqnarray}
                &&-\chi = \int \frac{\rho_{\bm J}(s)}{t(-\chi) - s} {\rm{d}}s, \\
                &&t(-\chi) = \frac{\rho_{\rm{CA}} - 1}{\chi}, \\
                &&t'(-\chi) = - \left[ \int \frac{\rho_{\bm J}(s)}{\left( t(-\chi) - s \right)^2} {\rm{d}}s \right]^{-1}.
            \end{eqnarray}
        \end{claim}
        
        The definition of the residual sum of squares $\overline{\rm{RSS}}$ is given in \eqref{eq:RSS}, while it is unrealistic to calculated in practice.
        In a single hypothesis testing, we estimate the value of $\overline{\rm{RSS}}$ in this way: 
        \begin{equation}
            \overline{\rm{RSS}} = \frac{1}{M}  \left\| \bm{y} - \bm{A} {\hat {\bm x}}^{\rm {LASSO}} \right\|_2^2.
        \end{equation}
    	
	\subsection{Proof of Claim \ref{claim:distribution}}
	\label{subsec:distribution}
    	
    	We evaluate the free energy density corresponding to the \ac{lasso} Hamiltonian $H({\bm{x}}) = {\left\| {\bm y} - {\bm A}{\bm x} \right\|_2^2/2} + \lambda {\left\|{\bm x} \right\|_1}$ at a zero-temperature limit: 
    	\begin{equation}
    	    \label{eq:FED}
    	    f(\lambda) \buildrel \textstyle. \over = - \mathop {\lim }\limits_{\beta  \to \infty } \mathop {\lim }\limits_{N \to \infty } \frac{1}{\beta N} {\mathbb{E}}_{{\bm A}, {\bm \xi}} \left[ \ln Z\left( {{\bm y}, {\bm A}; \lambda, \beta} \right) \right],
    	\end{equation}
    	where $\beta$ is the inverse temperature and $Z$ is the partition function: 
    	\begin{equation}
    	    Z\left( {{\bm y}, {\bm A}; \lambda, \beta} \right) = \int{ \exp{ \left( -\frac{\beta}{2} {\left\| {\bm y} - {\bm A}{\bm x} \right\|_2^2} - \beta \lambda {\left\|{\bm x} \right\|_1} \right)} {\rm{d}} {\bm{x}} }.
    	\end{equation}
    	In the zero-temperature limit $\beta \to \infty$, the Boltzmann distribution ${\rm{e}}^{-\beta H({\bm{x}})} / Z$ is dominated by the configurations of the \ac{lasso} solution.
    	Hence, one can evaluate how the \ac{lasso} estimator depends on ${\bm{x}}_0$, ${\bm A}$, ${\bm \xi}$ by analyzing the macroscopic behavior of the typical free energy density \eqref{eq:FED} using statistical mechanics.
    	
    	Based on the replica method and the \ac{rs} ansatz, the following is claimed.
    	\begin{claim}
    	    \label{clm:FED}
    	    When the steering matrix $\bm A$ is row-orthogonal, the free energy density \eqref{eq:FED} can be evaluated: 
    	    \begin{eqnarray}
        	    \label{eq:FEDRM}
                f &=& \mathop {\rm{extr}} \limits_{ Q, \hat Q, \chi, \hat \chi, m, \hat m } \left\{ G' (-\chi; {\bm{J}}) \left( Q - 2m + \rho - \frac{\chi}{2} \sigma^2 \right) + \frac{\gamma}{2}\sigma^2 - {\hat QQ} + {\hat \chi \chi} + 2 \hat mm \right. \nonumber \\ 
                && \left. + \mathop {\lim }\limits_{N \to \infty } \frac{1}{N} \sum\limits_{i = 1}^{N} {\int {\mathop {\min }\limits_{{x_i}} \left[ { \frac{{\hat Q}}{2}\left| {x_i} \right|^2 - {\rm{Re}} \left( \left( {\hat m{x_{0,i}} + \sqrt {2 \hat \chi } {z_i}} \right)^*{x_i} \right) + \lambda \left| {x_i} \right|} \right]{\rm{D}}{z_i}} }  \right\},
            \end{eqnarray}
        	where ${\rm{extr}}_X\{F(X)\}$ denotes extremization of a function $F(X)$ with respect to $X$ and $G^\prime (x; {\bm{J}})$ is the derivative of $G(x; {\bm{J}})$ with respect to $x$.
        	We define $\int{(\ldots) {\rm{D}}z}$, ${\bm{J}}$ and $G(x)$ as follows: 
    	    \begin{equation}
    	        \int{(\ldots){\rm{D}}z} \buildrel \textstyle. \over = \int{(\ldots) \frac{\exp(- \left| z \right|^2)}{\pi} {\rm{d}}z},
    	    \end{equation}
    	    \begin{equation}
    	        {\bm{J}} \buildrel \textstyle. \over = {\bm{A}}^H {\bm{A}},
    	    \end{equation}
    	    \begin{equation}
    	        \label{eq:G}
    	        G(x; {\bm{J}}) \buildrel \textstyle. \over = \mathop {\rm{extr}} \limits_{t} \left[ - \int{\rho_{\bm J}(s) \ln \left| t - s \right| {\rm{d}}s} + t x \right] - \ln \left| x \right| - 1,
    	    \end{equation}
    	    where $z$ is a complex number and $\rho_{\bm J}(s)$ is the asymptotic eigenvalue distribution of $\bm J$.
    	    The derivative of the function $G(x; {\bm{J}})$ has the following form:
    	    \begin{equation}
    	        \label{eq:defGprime}
    	        G^\prime (x; {\bm{J}}) =  t(x) - \frac{1}{x} ,
    	    \end{equation}
    	    where $t(x)$ is implicitly determined by the extreme value condition of \eqref{eq:G}:
    	    \begin{equation}
    	        \label{eq:Stieltjes_transform}
    	        x = \int{\frac{\rho_{\bm J}(s)}{t(x) - s} {\rm{d}} s}.
    	    \end{equation}
    	    \begin{proof}
                The proof is given in Appendix \ref{appendix:derivation}.
            \end{proof}
    	\end{claim}
    	
	    We can then obtain the possible distribution of LASSO solution by the free energy density. 
	    
        On one hand, from the extreme value condition of the free energy density \eqref{eq:FEDRM} of variable $\hat m$, that is $\frac{\partial f}{\partial \hat m} = 0$, we have
	    \begin{equation}
	        \label{eq:extrm}
	        m = \mathop {\lim }\limits_{N \to \infty } \frac{1}{2N} \sum\limits_{i = 1}^{N} {\int {\rm{Re}} \left( x_{0, i}^* {\hat x_i} \right) {\rm{D}}{z_i} },
	    \end{equation}
	    where ${\hat x_i}$ satisfies
	    \begin{equation}
	        \label{eq:x_i1}
	        {\hat x_i} = \mathop {\arg \min }\limits_{{x_i}} \left[ { \frac{{\hat Q}}{2}\left| {x_i} \right|^2 - {\rm{Re}} \left( \left( {\hat m{x_{0,i}} + \sqrt {2 \hat \chi } {z_i}} \right)^*{x_i} \right) + \lambda \left| {x_i} \right|} \right].
	    \end{equation}
	    The above optimization problem has its analytical solution, given by
	    \begin{equation}
	        \label{eq:x_i2}
	        {\hat x_i} = {\rm{ST}}_{\lambda, \hat Q} \left( h_i \right),
	    \end{equation}
	    where $h_i$ is given in \eqref{eq:h_i}.
	    
	    On the other hand, from definition of the macroscopic physical observable $m$, that is $m = \frac{1}{2N} {\rm{Re}} \left( {\bm{x}}_0^H {\bm{x}}_a \right)$, we have
	    \begin{equation}
	        \label{eq:rsm}
	        m = \mathop {\lim }\limits_{N \to \infty } \frac{1}{2N} {\mathbb{E}}_{{\bm A}, {\bm \xi}} \left[ {\rm{Re}} \left( {\bm x}_0^H {\hat {\bm x}}^{\rm {LASSO}} \right) \right].
	    \end{equation}
	    Comparing \eqref{eq:extrm} with \eqref{eq:rsm}, the distribution of the LASSO solution $\hat {\bm{x}}^{\rm{LASSO}}$ can be inferred as follow:
	    \begin{equation}
	        \hat x_i^{\rm{LASSO}} = \hat x_i = {\rm{ST}}_{\lambda, \hat Q} \left( h_i \right).
	    \end{equation}

	    
    \subsection{Proof of Claim \ref{claim:debiased}}
    \label{subsec:debiased}
	    
	    Denote by ${\bm h} = \left[ h_1, h_2, \ldots, h_N \right]^T$, where $h_i$ is defined in \eqref{eq:h_i}.
	    In the \ac{tap} analysis, ${\bm h}$ is called local field.
	    Denote by $\left \langle {\bm x} \right \rangle$ the average of ${\bm x}$ taken by the Boltzmann distribution ${\rm{e}}^{- \beta H({\bm{x}})} / Z$, which is given by
	    \begin{equation}
	    \label{eq:Boltzmann_average}
	        \left \langle {\bm x} \right \rangle = \frac{\int{ {\bm x} \exp{ \left( -\frac{\beta}{2} {\left\| {\bm y} - {\bm A}{\bm x} \right\|_2^2} - \beta \lambda {\left\|{\bm x} \right\|_1} \right)} {\rm{d}} {\bm{x}} }}{\int{ \exp{ \left( -\frac{\beta}{2} {\left\| {\bm y} - {\bm A}{\bm x} \right\|_2^2} - \beta \lambda {\left\|{\bm x} \right\|_1} \right)} {\rm{d}} {\bm{x}} }}.
	    \end{equation}
	    In the zero-temperature limit $\beta \to \infty$, the average $\left \langle {\bm x} \right \rangle$ is equal to the LASSO solution ${\hat {\bm x}}^{\rm {LASSO}}$.
	    \ac{tap} approach is to obtain the mean field equation, which describes the relation between the local field ${\bm h}$ and the average $\left \langle {\bm x} \right \rangle$.
	    
	    Consider the following alternative Gibbs free energy \cite{takahashi2018statistical}: 
	    \begin{equation}
	    \label{eq:GmQ}
	        Gi({\bm m}, Q) = \mathop {\rm{extr}} \limits_{ {\bm h}, \Lambda } \left\{ {\rm{Re}} \left( {\bm h}^H {\bm m} \right) - N \Lambda Q - \frac{1}{\beta} \ln \int{ {\rm{e}}^{ - \frac{\beta}{2} \left\| {\bm y} - {\bm A} {\bm x} \right\|_2^2 + \beta {\rm{Re}} \left( {\bm h}^H {\bm x} \right) -\frac{\beta}{2} \Lambda \left\| {\bm x} \right\|_2^2 - \beta \lambda \left\| {\bm x} \right\|_1 } {\rm{d}}{\bm x} } \right\}.
	    \end{equation}
	    From the extreme value condition on $\bm h$, $\bm m$ $\Lambda$ and $Q$ of $G({\bm m}, Q)$, one can conclude that
	    \begin{equation}
	    \label{eq:def_mQ}
	        \left \langle {\bm x} \right \rangle, \left \langle \left\| {\bm x} \right\|_2^2 \right \rangle / (2N) = \mathop {\arg \min} \limits_{ {\bm m}, Q } G({\bm m}, Q).
	    \end{equation}
	    We will evaluate it by expectation consistent approximate inference \cite{opper2005expectation}.
	    \begin{claim}
	        The alternative Gibbs free energy \eqref{eq:GmQ} possesses the following expression under the limit $\beta \to \infty$:
	        \begin{eqnarray}
                \label{eq:GmQ2}
    	        Gi({\bm m}, Q) &=& \mathop {\rm{extr}} \limits_{ {\bm h}, \Lambda } \left\{ {\rm{Re}} \left( {\bm h}^H {\bm m} \right) - N \Lambda Q - {\sum \limits_{i = 1}^N \frac{\left( \left| h_i \right| - \lambda \right)^2}{2 \Lambda} \cdot \Theta \left( \left| h_i \right| - \lambda \right) } \right\} \nonumber \\
    	        &&- \frac{N}{\beta} G(-\chi; {\bm{J}}) + \frac{1}{2} \left\| {\bm y} - {\bm A} {\bm m} \right\|_2^2,
    	    \end{eqnarray}
    	    where $\chi = \beta(Q-q)$ and $q = \sum \limits_{i} m_i^2 /N$.
    	    \begin{proof}
    	        The expectation consistent inference provides the following approximation:
        	    \begin{equation}
        	    \label{eq:ec}
        	        Gi({\bm m}, Q) \simeq \phi_{ada}({\bm m}, Q) = \tilde \phi({\bm m}, Q; l = 0) + \phi^G({\bm m}, Q; l = 1) - \phi^G({\bm m}, Q; l = 0),
        	    \end{equation}
        	    where 
        	    \begin{equation}
        	        \label{eq:tilde_phi}
        	        \tilde \phi({\bm m}, Q; l) = \mathop {\rm{extr}} \limits_{ {\bm h}, \Lambda } \left\{ {\rm{Re}} \left( {\bm h}^H {\bm m} \right) - N \Lambda Q - \frac{1}{\beta} \ln \int{ {\rm{e}}^{ - \frac{\beta l}{2} \left\| {\bm y} - {\bm A} {\bm x} \right\|_2^2 + \beta {\rm{Re}} \left( {\bm h}^H {\bm x} \right) -\frac{\beta}{2} \Lambda \left\| {\bm x} \right\|_2^2 - \beta \lambda \left\| {\bm x} \right\|_1 } {\rm{d}}{\bm x} } \right\},
        	    \end{equation}
        	    and
        	    \begin{equation}
        	        \label{eq:phiG}
        	        \phi^G({\bm m}, Q; l) = \mathop {\rm{extr}} \limits_{ {\bm h}, \Lambda } \left\{ {\rm{Re}} \left( {\bm h}^H {\bm m} \right) - N \Lambda Q - \frac{1}{\beta} \ln \int{ {\rm{e}}^{ - \frac{\beta l}{2} \left\| {\bm y} - {\bm A} {\bm x} \right\|_2^2 + \beta {\rm{Re}} \left( {\bm h}^H {\bm x} \right) -\frac{\beta}{2} \Lambda \left\| {\bm x} \right\|_2^2 } {\rm{d}}{\bm x} } \right\}.
        	    \end{equation}
        	    With such approximation, $\tilde \phi({\bm m}, Q; l = 1)$, $\phi^G({\bm m}, Q; l = 1)$ and $\phi^G({\bm m}, Q; l = 0)$ can be easily calculated as follow: 
        	    \small
        	    \begin{eqnarray}
        	        \label{eq:tilde_phi0}
        	        &&\tilde \phi({\bm m}, Q; l = 0) = \mathop {\rm{extr}} \limits_{ {\bm h}, \Lambda } \left\{ {\rm{Re}} \left( {\bm h}^H {\bm m} \right) - N \Lambda Q - {\sum \limits_{i = 1}^N \frac{\left( \left| h_i \right| - \lambda \right)^2}{2 \Lambda} \cdot \Theta \left( \left| h_i \right| - \lambda \right) } \right\}, \\
        	        \label{eq:phiG1}
        	        &&\phi^G({\bm m}, Q; l = 1) = - \frac{N}{\beta} G(-\chi; {\bm{J}}) + \frac{1}{2} \left\| {\bm y} - {\bm A} {\bm m} \right\|_2^2 + \frac{N}{\beta} \ln \frac{\beta}{2 \pi} - \frac{N}{\beta} \ln \left| - \chi \right| - \frac{N}{\beta}, \\
        	        \label{eq:phiG0}
        	        &&\phi^G({\bm m}, Q; l = 0) = \frac{N}{\beta} \ln \frac{\beta}{2 \pi} - \frac{N}{\beta} \ln \left| - \chi \right| - \frac{N}{\beta},
        	    \end{eqnarray}
        	    \normalsize
        	    where \eqref{eq:tilde_phi0} is calculated under the limit $\beta \to \infty$.
        	    Therefore, 
        	    \begin{eqnarray}
                \label{eq:phiada}
        	        \phi_{ada}({\bm m}, Q) &=& \mathop {\rm{extr}} \limits_{ {\bm h}, \Lambda } \left\{ {\rm{Re}} \left( {\bm h}^H {\bm m} \right) - N \Lambda Q - {\sum \limits_{i = 1}^N \frac{\left( \left| h_i \right| - \lambda \right)^2}{2 \Lambda} \cdot \Theta \left( \left| h_i \right| - \lambda \right) } \right\} \nonumber \\
        	        &&- \frac{N}{\beta} G(-\chi; {\bm{J}}) + \frac{1}{2} \left\| {\bm y} - {\bm A} {\bm m} \right\|_2^2.
        	    \end{eqnarray}
    	    \end{proof}
	    \end{claim}
	    
	    
        Taking the limit $\beta \to \infty$ and $N \to \infty$, we can get the mean field equation by the extreme value condition on ${\bm h}$, $\Lambda$, ${\bm m}$, $Q$ of \eqref{eq:phiada} and linear response argument \cite{tanaka2002statistical, opper2001advanced}, given by 
	    \begin{eqnarray}
	        \label{eq:bm_h}
	        {\bm h} &=& \Lambda {\bm m} + {\bm A}^H ({\bm y} - {\bm A} {\bm m}), \\
	        m_i &=& \frac{h_i}{\left| h_i \right|} \cdot \frac{\left| h_i \right| - \lambda}{\Lambda} \cdot \Theta \left( \left| h_i \right| - \lambda \right), \\
	        \Lambda &=& G'(-\chi; {\bm{J}}), \\
	        \label{eq:getchi}
	        \chi &=& \frac{1}{2 N \Lambda} {\sum \limits_{i = 1}^N \left[ \left( 2 - \frac{\lambda}{\left| h_i \right|} \right) \cdot  \Theta \left( \left| h_i \right| - \lambda \right) \right] },
	    \end{eqnarray}
	    where \eqref{eq:getchi}, derived in Appendix \ref{appendix:LRA}, is obtained by linear response argument.
	    Denote by $\rho_{\rm{CA}}$ the complex active component density of the LASSO solution as follow:
	    \begin{equation}
	        \rho_{\rm{CA}} = \frac{1}{2N} {\sum \limits_{i = 1}^N \left[ \left( 2 - \frac{\lambda}{\left| h_i \right|} \right) \cdot  \Theta \left( \left| h_i \right| - \lambda \right) \right] },
	    \end{equation}
	    which is the same as \eqref{eq:alg_rho_CA}.
	    According to \eqref{eq:bm_h}, one can obtain the debiased estimator ${\hat {\bm x}}^{\rm {d}}$ as follow:
	    \begin{equation}
	        {\hat {\bm x}}^{\rm {d}} = {\hat {\bm x}}^{\rm {LASSO}} + \frac{1}{\Lambda} {\bm A}^H ({\bm y} - {\bm A} {\hat {\bm x}}^{\rm {LASSO}}),
	    \end{equation}
	    where $\Lambda = G'(-\chi; {\bm{J}})$ and $\chi$ can be obtained by solving
	    \begin{equation}
	        \label{eq:eqchi}
	        \frac{\rho_{\rm{CA}}}{\chi} = G'(-\chi; {\bm{J}}).
	    \end{equation}
	    Then we get the following equations according to the extreme condition in function $G(-\chi; {\bm{J}})$ \eqref{eq:Stieltjes_transform} and \eqref{eq:eqchi}: 
	    \begin{eqnarray}
	        &&\Lambda = G'(-\chi; {\bm{J}}) = \frac{\rho_{\rm{CA}}}{\chi} = t(- \chi) + \frac{1}{\chi},\\
            \label{eq:tchi}
	        &&- \chi = \frac{1 - \gamma}{t(-\chi)} +  \frac{\gamma}{t(-\chi) - 1}, 
	    \end{eqnarray}
	    which lead to \eqref{eq:Lambda} and \eqref{eq:eq_t}. 
	    
    \subsection{Proof of Claim \ref{claim:variance}}
    \label{subsec:variance}
        
        We first define some macroscopic observables. 
        \begin{claim}
            The following relationships between the free energy density, regularization term, and residual sum of squares hold:
            \begin{eqnarray}
                &&f = \frac{\gamma}{2} \overline{\rm{RSS}} + \bar r, \\
                \label{eq:r}
                &&\bar r = \mathbb{E}_{\bm{A}, \bm{\xi}} \left[ \left\langle \frac{\lambda}{N} {\sum \limits_{i = 1}^N \left| x_i \right|}  \right\rangle \right] = 2 \hat \chi \chi + 2 \hat m m - 2 \hat Q Q, \\
                \label{eq:RSS}
                &&\overline{\rm{RSS}} = \mathbb{E}_{\bm{A}, \bm{\xi}} \left[ \rm{RSS} \right] = \mathbb{E}_{\bm{A}, \bm{\xi}} \left[ \left\langle \frac{1}{M}  \left\| \bm{y} - \bm{A} \bm{x} \right\|_2^2  \right\rangle \right] \\
                \label{eq:expRSS}
                && = \frac{2}{\gamma} \left[ G' (-\chi; {\bm{J}}) (Q - 2m + \rho - \frac{\chi}{2} \sigma^2) + \frac{\gamma}{2}\sigma^2 - \hat \chi \chi \right],
            \end{eqnarray}
            where $\bar r$ and $\overline{\rm{RSS}}$ represent the per-element average of the regulation term and residual sum of squares, respectively.
            \begin{proof}
                Consider the extreme value condition of the free energy density $f$ \eqref{eq:FEDRM} on $\hat Q$, $\hat m$ and $\hat \chi$, given by
        	     \begin{eqnarray}
        	         &&\frac{\partial f}{\partial \hat Q} = - Q + \mathop {\lim }\limits_{N \to \infty } \frac{1}{N} \sum\limits_{i = 1}^{N} {\int   \frac{1}{2} \left| {\hat x_i} \right|^2  {\rm{D}}{z_i} } = 0, \\
        	         &&\frac{\partial f}{\partial \hat m} = 2m - \mathop {\lim }\limits_{N \to \infty } \frac{1}{N} \sum\limits_{i = 1}^{N} {\int   {\rm{Re}} \left( x_{0,i}^* {\hat x_i} \right) {\rm{D}}{z_i} } = 0, \\
        	         &&\frac{\partial f}{\partial \hat \chi} = \chi - \mathop {\lim }\limits_{N \to \infty } \frac{1}{N} \sum\limits_{i = 1}^{N} {\int {\rm{Re}} \left( \frac{1}{\sqrt{2 \hat \chi}} z_i^* {\hat x_i} \right)  {\rm{D}}{z_i} } = 0,
        	     \end{eqnarray}
        	     where $\hat x_i$ is the same with \eqref{eq:x_i1} and \eqref{eq:x_i2}.
        	     The above equations note that
        	     \begin{eqnarray}
        	         && \hat Q Q = \mathop {\lim }\limits_{N \to \infty } \frac{1}{N} \sum\limits_{i = 1}^{N} {\int   \frac{\hat Q}{2} \left| {\hat x_i} \right|^2  {\rm{D}}{z_i} }, \\
        	         && 2 \hat m m = \mathop {\lim }\limits_{N \to \infty } \frac{1}{N} \sum\limits_{i = 1}^{N} {\int {\rm{Re}} \left( \left( \hat m x_{0,i} \right)^* {\hat x_i} \right) {\rm{D}}{z_i} }, \\
        	         && 2 \hat \chi \chi = \mathop {\lim }\limits_{N \to \infty } \frac{1}{N} \sum\limits_{i = 1}^{N} {\int {\rm{Re}} \left( \left( \sqrt{2 \hat \chi} z_i \right)^* {\hat x_i} \right)  {\rm{D}}{z_i} }.
        	     \end{eqnarray}
        	     Therefore, one can obtain that
        	     \begin{eqnarray}
        	         && \mathop {\lim }\limits_{N \to \infty } \frac{1}{N} \sum\limits_{i = 1}^{N} {\int {\mathop {\min }\limits_{{x_i}} \left[ { \frac{{\hat Q}}{2}\left| {x_i} \right|^2 - {\rm{Re}} \left( \left( {\hat m{x_{0,i}} + \sqrt {2 \hat \chi } {z_i}} \right)^*{x_i} \right) + \lambda \left| {x_i} \right|} \right]{\rm{D}}{z_i}} } \nonumber \\
        	         &=& \hat Q Q - 2 \hat \chi \chi - 2 \hat m m + \bar r.
        	     \end{eqnarray}
        	     On the other hand, with the close-form expression of $\hat x_i$, we can get that
        	     \begin{eqnarray}
        	         &&\mathop {\lim }\limits_{N \to \infty } \frac{1}{N} \sum\limits_{i = 1}^{N} {\int {\mathop {\min }\limits_{{x_i}} \left[ { \frac{{\hat Q}}{2}\left| {x_i} \right|^2 - {\rm{Re}} \left( \left( {\hat m{x_{0,i}} + \sqrt {2 \hat \chi } {z_i}} \right)^*{x_i} \right) + \lambda \left| {x_i} \right|} \right]{\rm{D}}{z_i}} } \nonumber \\
        	         &=& \mathop {\lim }\limits_{N \to \infty } \frac{1}{N} \sum\limits_{i = 1}^{N} {\int -  \frac{\hat Q}{2} \left| {\hat x_i} \right|^2  {\rm{D}}{z_i} } = - \hat Q Q.
        	     \end{eqnarray}
        	     The expression of the regulation term $\bar r$ \eqref{eq:r} can then be derived, and the same with $\overline{\rm{RSS}}$ \eqref{eq:expRSS}.
            \end{proof}
        \end{claim}
        
        Recall the extreme value condition of the free energy density $f$ \eqref{eq:FEDRM} on $Q$, $m$ and $\chi$, given by
        \begin{eqnarray}
            &&\frac{\partial f}{\partial Q} = G'(-\chi; {\bm J}) - \hat Q, \\
            &&\frac{\partial f}{\partial m} = 2 G'(-\chi; {\bm J}) - 2 \hat m, \\
            \label{eq:pfpchi}
            &&\frac{\partial f}{\partial \chi} = -G''(-\chi; {\bm J}) \left( Q - 2m + \rho - \frac{\chi}{2} \sigma^2 \right) - \frac{\sigma^2}{2} G'(-\chi; {\bm J}) + \hat \chi = 0.
        \end{eqnarray}
        Together with the expression of $\overline{\rm{RSS}}$ \eqref{eq:expRSS}, the expression of $\hat \chi$ \eqref{eq:eqhatchi} can be obtained.
        The derivatives of function $G(-\chi; {\bm J})$ \eqref{eq:Gprimechi} and \eqref{eq:Gwprimechi} follow the derivation of \eqref{eq:defGprime}.
        Derivatives act on \eqref{eq:Stieltjes_transform} leads to the expressions of $t(-\chi)$ and $t'(-\chi)$.

    \section{Numerical Experiments}
    \label{sec:result}
    
    In this section, we mainly provide the numerical simulations to exam the following capabilities. 
    \begin{enumerate}
        \item Gaussianity of $\bm{w}$ in the case of row-orthogonal matrix design, where $\bm{w} = \hat{\bm{x}}^{\rm{d}} - {\bm{x}}_0$ denotes the difference between the debiased \ac{lasso} estimator and the original signal ${\bm{x}}_0$. 
        \item Accuracy of estimating the variance $\sigma_w^2$ in both the case of row-orthogonal and Gaussian matrix design.
        \item Detection performance of debiased \ac{lasso} detector.
    \end{enumerate}
        
        \subsection{Settings}
        \label{subsec:settings}
        
            In all the numerical experiments, we artificially generate the original signal ${\bm{x}}_0$, observation matrix ${\bm{A}}$ and the noise ${\bm{\xi}}$. 
            The original signal ${\bm{x}}_0$ is generated from the Bernoulli-Gaussian distribution: $p_x = (1 - 2 \rho) \delta(x) + \frac{2 \rho}{\pi \sigma_x^2} {\rm{e}}^{-|x|^2 / \sigma_x^2}$. 
            For the observation matrix ${\bm{A}}$, we separately consider Gaussian design and row-orthogonal design. 
            The former is achieved by setting all the entries of $\bm A$ i.i.d. complex Gaussian variables: $A_{ij} \sim {\cal{CN}}(0, 1/N)$, and the latter is achieved by randomly selecting $M$ rows from a randomly generated $N \times N$ orthogonal matrix. 
            The entries of the noise ${\bm{\xi}}$ are i.i.d. complex Gaussian variables: $\xi_i \sim {\cal{CN}}(0, \sigma^2)$. 
            We consider the \ac{mf} definition of the \ac{snr} \cite{na2018tendsur}, such that
            \begin{equation}
                {\rm{SNR}} = \frac{\gamma \sigma_x^2}{\sigma^2}.
            \end{equation}
    
        \subsection{Gaussianity of $\bm{w}$ in the case of row-orthogonal matrix design}
        \label{subsec:Gaussianity}
        
            \begin{table}
                \caption{p-values of Kolmogorov-Smirnov test of the real and imaginary parts of non-zero entries and zero entries of ${\bm{w}}^{\rm{CROM}}$ and $\bm{w}^{\rm{CG}}$.}
                \begin{center}  
                    \begin{tabular}{ccccc} 
                        \hline 
                        \textbf{p-values} & ${\bm{w}}^{\rm{CROM}}$ $({\cal{H}}_1)$ & ${\bm{w}}^{\rm{CROM}}$ $({\cal{H}}_0)$ & $\bm{w}^{\rm{CG}}$ $({\cal{H}}_1)$ & $\bm{w}^{\rm{CG}}$ $({\cal{H}}_0)$ \\
                        \hline 
                        \textbf{real part} & $0.7563$ & $0.6702$ & $1.731 \times 10^{-78}$ & $0$ \\
                        \textbf{imaginary part} & $0.8259$ & $0.6090$ & $1.729 \times 10^{-70}$ & $0$ \\
                        \hline 
                    \end{tabular}  
                \end{center}
                \label{tab:KStest} 
            \end{table}
            
            \begin{figure}
                \centering
                \includegraphics[width=8cm]{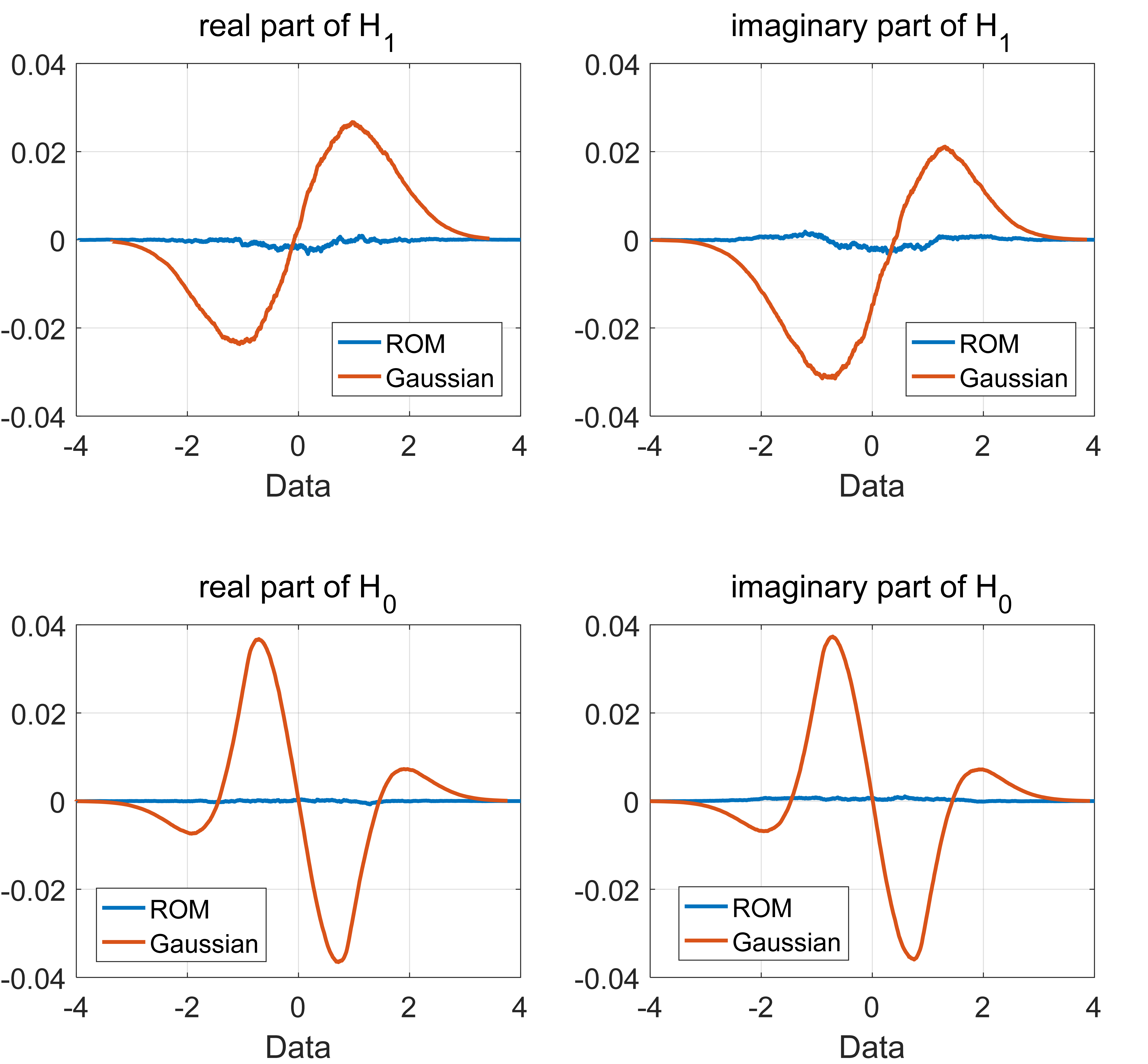}
                \caption{The difference between the Gaussian distribution and the empirical CDF of the real and imaginary parts of non-zero entries and zero entries of ${\bm{w}}^{\rm{CROM}}$ and ${\bm{w}}^{\rm{CG}}$.}
                \label{Fig:Gaussianity}
            \end{figure}
            
            We investigate the Gaussianity of ${\bm{w}}$ in the case of row-orthogonal matrix design by comparing the \ac{ecdf} of its real and imaginary part with Gaussian distribution. 
            The comparisons are performed for the complex row-orthogonal debiased estimator $\hat{\bm{x}}^{\rm{d, CROM}}$ (with $\Lambda_{\rm{CROM}}$ given by \eqref{eq:Lambda_CROM}) and the complex Gaussian debiased estimator $\hat{\bm{x}}^{\rm{d, CG}}$ (with $\Lambda_{\rm{CG}}$ given by \eqref{eq:Lambda_CG}). 
            Denote by ${\bm{w}}^{\rm{CROM}} = \hat{\bm{x}}^{\rm{d, CROM}} - {\bm{x}}_0$ and ${\bm{w}}^{\rm{CG}} = \hat{\bm{x}}^{\rm{d, CG}} - {\bm{x}}_0$. 
            Let ${\tilde{\bm{w}}}^{\rm{CROM}} = \sqrt{2} {\bm{w}}^{\rm{CROM}} / \sigma_{w, {\rm{CROM}}}^2$, ${\tilde{\bm{w}}}^{\rm{CROM}} = \sqrt{2} {\bm{w}}^{\rm{CG}} / \sigma_{w, {\rm{CG}}}^2$. 
            Consequently, the \ac{ecdf} of the real and imaginary part of these two vectors, denoted by $F_{{\rm{CROM}}, r}(x)$, $F_{{\rm{CROM}}, i}(x)$, $F_{{\rm{CG}}, r}(x)$ and $F_{{\rm{CG}}, i}(x)$ respectively, are considered to converge to standard Gaussian distribution $\Phi(x)$ weakly. 
            We divide ${\tilde{\bm{w}}}^{\rm{CROM}}$ and ${\tilde{\bm{w}}}^{\rm{CG}}$ into 4 parts: real and imaginary part of non-zero entries and zero entries, and verify their Gaussianity by demonstrating the differences such as $\Phi(x) - F_{{\rm{CROM}}, {\cal{H}}_1, r}(x)$ respectively. 
            
            The observation matrix $\bm A$ is partial Fourier with a size of $M = 768$ and $N = 1024$.
            The variance $\sigma_x^2$ of the non-zero entries in ${\bm x}_0$ is set to be $1$ and the regularization parameter $\lambda$ of \ac{lasso} is $0.1$. 
            We set the signal density to be $2\rho = 0.1$ and \ac{snr} to be $5$dB. 
            The empirical laws are obtained by $10^3$ Monte-Carol trials. 
            
            Fig. \ref{Fig:Gaussianity} shows the difference between Gaussian distribution $\Phi(x)$ and the \ac{ecdf} of each part of ${\bm{w}}^{\rm{CROM}}$ and ${\bm{w}}^{\rm{CG}}$. 
            The results shows that in the case of partial Fourier matrix design, the \ac{ecdf} of ${\bm{w}}^{\rm{CROM}}$ is very close to the Gaussian distribution, while the \ac{ecdf} of ${\bm{w}}^{\rm{CG}}$ has a significant difference from the Gaussian distribution. 
            We also employ \ac{ks} test \cite{massey1951kolmogorov} on them, with the p-values shown in Table \ref{tab:KStest}. 
            The \ac{ks} test compares the \ac{ecdf} of the input samples with a certain distribution (here we set it to be Gaussian distribution) and the larger of p-values means the higher probability that the samples come from the given distribution.
            It is obvious that for each part of ${\bm{w}}^{\rm{CROM}}$, \ac{ks} test verifies that the \ac{ecdf} well meets the expected theoretical distribution while the other one indicates the opposite result. 
            Our verification of the Gaussianity of ${\bm{w}}^{\rm{CROM}}$ in turn verifies the correctness of the debiased coefficient $\Lambda_{\rm{CROM}}$ in the case of row-orthogonal matrix design. 
            
        \subsection{Accuracy of estimating $\sigma_w^2$}
        \label{subsec:Accuracy}
        
            \begin{figure}
                \centering
                \includegraphics[width=8cm]{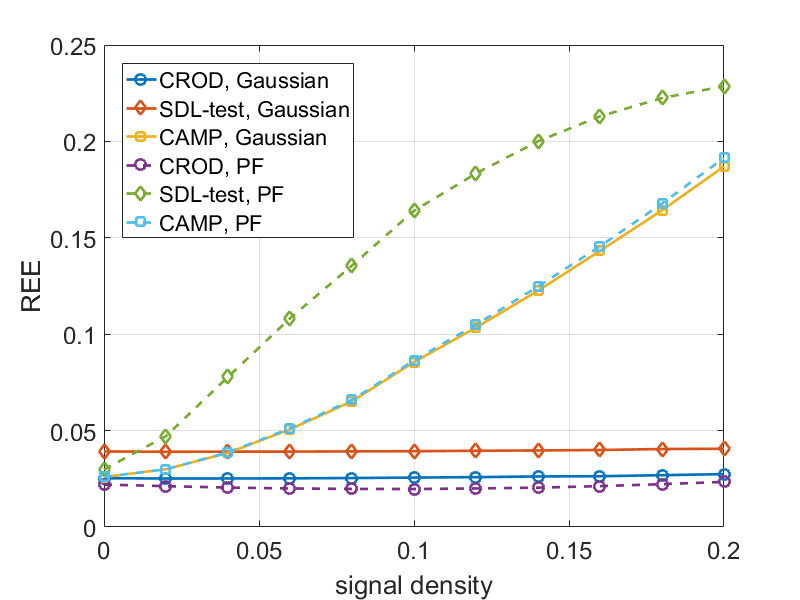}
                \caption{REE of CROD, CAMP and SDL-test under the Gaussian and partial Fourier design model.}
                \label{Fig:std_est_Gaussian_PF_rho}
            \end{figure}
            
            \begin{figure}
                \centering
                \includegraphics[width=8cm]{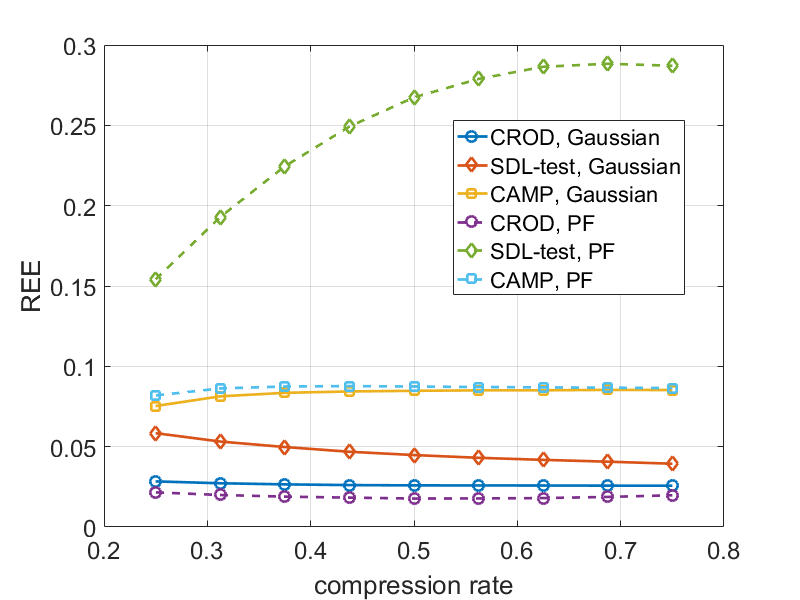}
                \caption{REE of CROD, CAMP and SDL-test under the Gaussian and partial Fourier design model.}
                \label{Fig:std_est_Gaussian_PF_gamma}
            \end{figure}

            In this section, we investigate the accuracy of the estimation of $\sigma_w^2$ by comparing the relative estimation error of different approaches, which is defined as follow
            \begin{equation}
                \label{eq:REE}
                {\rm{REE}} = \frac{\left| \hat \sigma_w - \sigma_w \right|}{\sigma_w},
            \end{equation}
            where $\hat \sigma_w$ is the estimated result and the ground truth $\sigma_w$ is obtained by 
            \begin{equation}
                \label{eq:gt_sigma_w}
                \sigma_w = \sqrt{\frac{1}{N} \sum \limits_{i = 1}^N \left| w_i \right|^2}.
            \end{equation}
            We compare ${\rm{REE}}$ of \ac{crod}, CAMP and SDL-test in the case of Gaussian matrix design and row-orthogonal matrix design, respectively. 
            In the following experiments, we set the length of ${\bm x}_0$ to be $N = 256$ and $\sigma_x^2$ to be $1$. 
            The variance of noise $\sigma^2 = 0.05$, corresponding to a \ac{snr} of $13$dB.
            Noting that the estimation approach of SDL-test in \cite{javanmard2014hypothesis} is not suitable for complex-valued data, we have modified it to the following form: 
            \begin{equation}
                \label{eq:variance_SDL_complex}
                \hat \sigma_w = \frac{\sqrt{\gamma}}{\sqrt{\ln 2} (\gamma - \rho_{CA})} {\rm{median}} \left( \left| {\bm y} - {\bm A} {\hat {\bm x}}^{\rm {LASSO}} \right| \right), 
            \end{equation}
            
            The results are obtained by $10^5$ Monte-Carol trials and presented in Fig. \ref{Fig:std_est_Gaussian_PF_rho} and Fig. \ref{Fig:std_est_Gaussian_PF_gamma}, in which the signal density and compression rate vary, respectively. 
            The results of the numerical experiments illustrate that the variance estimation method we use has the ability to consistently obtain high estimation accuracy under a variety of different conditions. 
            
		\subsection{Detection performance of debiased \ac{lasso} detector}
		\label{subsec:detection_performance}
		
			\begin{figure}
				\centering
				\includegraphics[width=8cm]{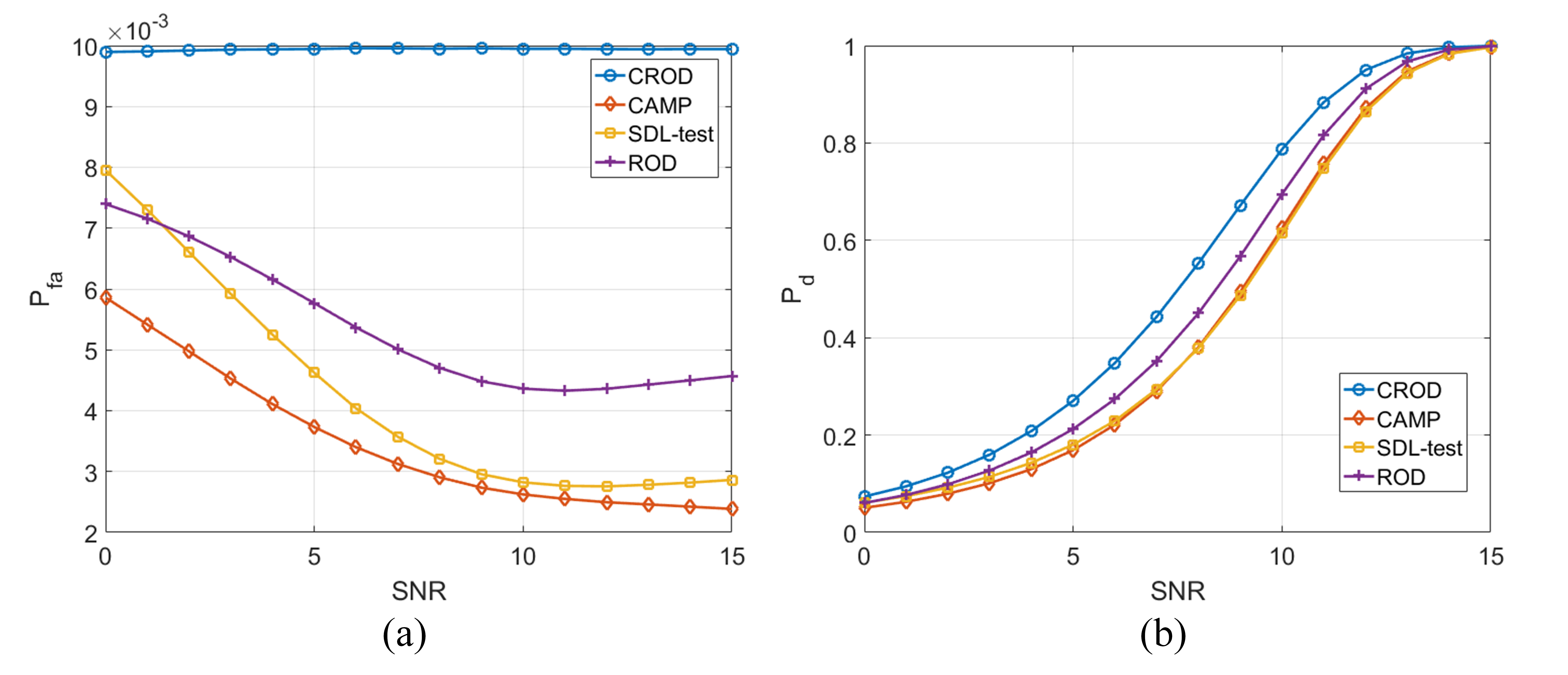}
				\caption{(a) Probability of false alarm and (b) detection of different detectors vary with SNR. Here $2\rho = 0.1$ and $\gamma = 0.5$.}
				\label{Fig:Pfa_Pd_PF_SNR}
			\end{figure}

			\begin{figure}
		       	\centering
		       	\includegraphics[width=8cm]{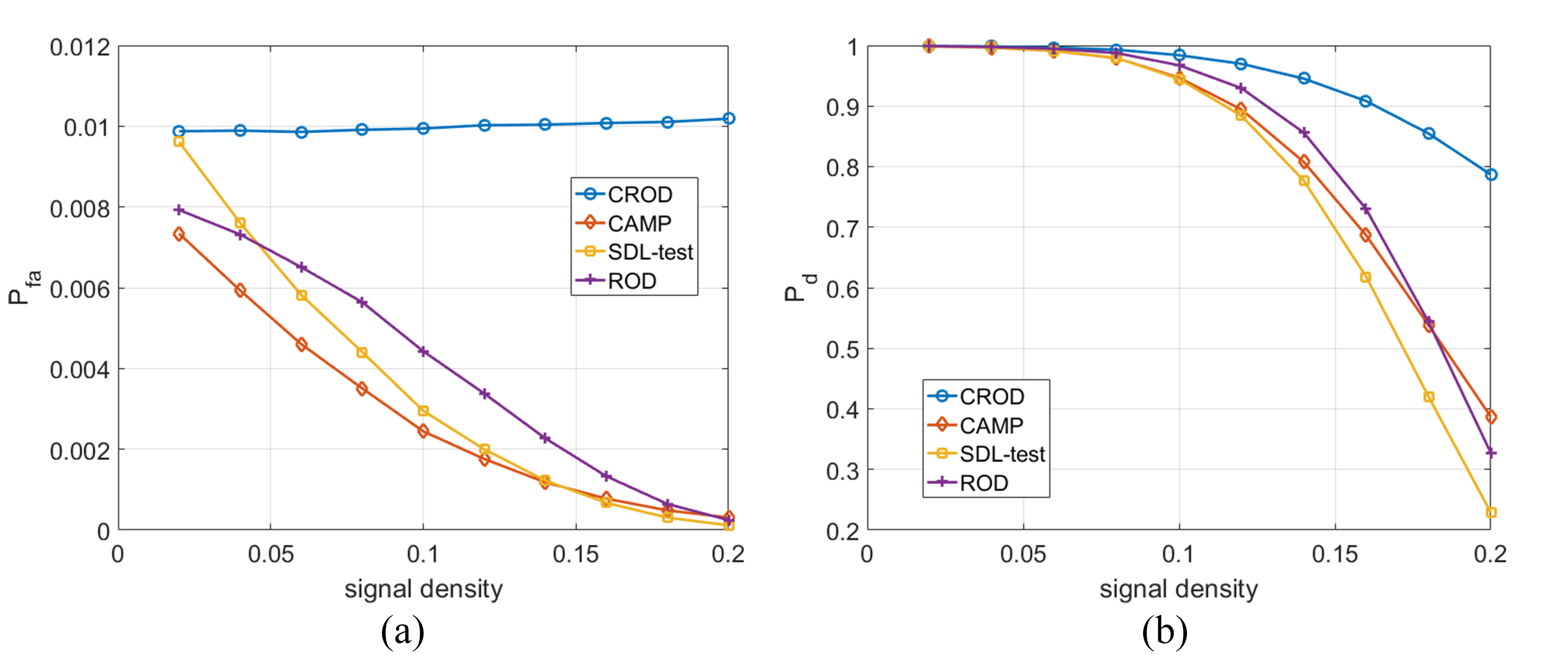}
		       	\caption{(a) Probability of false alarm and (b) detection of different detectors vary with signal density. Here ${\rm{SNR}} = 13$dB and $\gamma = 0.5$.}
		       	\label{Fig:Pfa_Pd_PF_p0}
			\end{figure}
			
			\begin{figure}
		       	\centering
		       	\includegraphics[width=8cm]{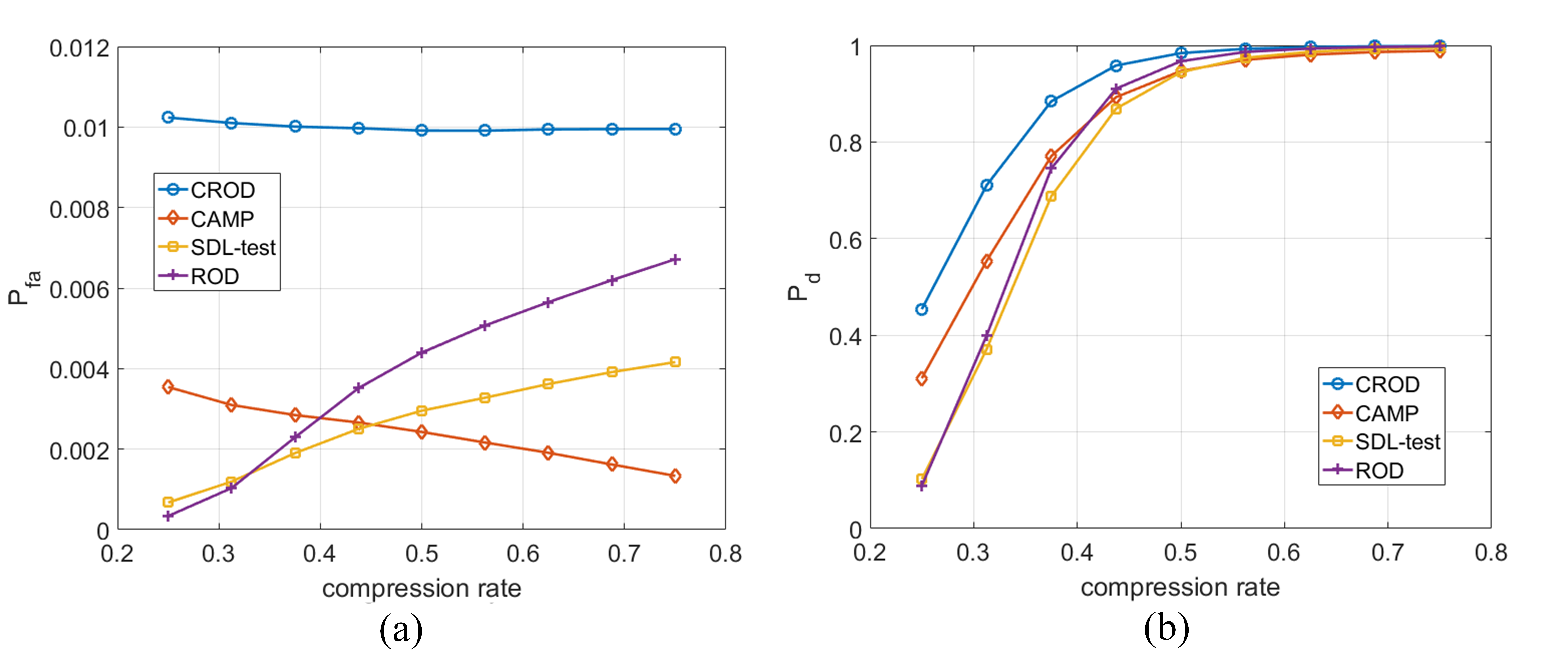}
		       	\caption{(a) Probability of false alarm and (b) detection of different detectors vary with compression rate. Here $2\rho = 0.1$ and ${\rm{SNR}} = 13$dB.}
		       	\label{Fig:Pfa_Pd_PF_gamma}
			\end{figure}
			
			We apply several detectors for solving the sub-Nyquist radar detection problem, in which the observation matrix $\bm A$ is partial Fourier, and examine their detection performance. 
			Debiased detectors listed in Table \ref{tab:detectors_comparison}, including \ac{crod}, CAMP, SDL-test and \ac{rod}, are compared. 
			In the following experiments, we set the length of ${\bm x}_0$ to $N = 256$ and $\sigma_x^2$ to $1$. 
			The probability of false alarm is set to $0.01$ and all the results are obtained by $10^5$ Monte-Carol trials. 
			In Fig. \ref{Fig:Pfa_Pd_PF_SNR}, \ref{Fig:Pfa_Pd_PF_p0} and \ref{Fig:Pfa_Pd_PF_gamma}, we demonstrate how the detection performance of these detectors varies with \ac{snr}, signal density, and compression rate, respectively. 
			The results suggest that \ac{crod} has the best ability to maintain the false alarm rate under multiple parameter variations, i.e., it can most accurately calculate the threshold (or p-value) for a given false alarm rate. 
    
    \section{Conclusion}
    \label{sec:conclusion}
    
        In this paper, we provide the design of debiased \ac{lasso} detector solving the detection problem of compressed sensing radar under the row-orthogonal design model. 
        The detection performance of the present detector is theoretically analyzed and proved to be better than \ac{lasso} detector. 
        We also compare the proposed approach with other debiased \ac{lasso} detectors, simulation results indicate that our approach can provide test statistic and threshold (or p-value) more accurately. 
        Such merit allows precise control of the false alarm rate, resulting in the higher detection rate than other \ac{cs} detectors.
    
    \appendix

    \section{Proof of Theorem \ref{theorem:lambda_unique}}
    \label{appendix:unique}

        Assume that there exists $\Lambda_2 \ne \Lambda_1$ such that the empirical law of ${\bm{w}}_2(N)$ converges to a Gaussian distribution with mean $\mu_2$ as $N \to \infty$. 
        According to Lemma \ref{lemma:solution_condition}, we have the following inequality
        \begin{equation}
            \label{eq:w_2i_w_1i}
            \left| w_{2, i} - w_{1, i} \right| \le \lambda \left| \frac{1}{\Lambda_2} - \frac{1}{\Lambda_1} \right| \buildrel \textstyle. \over = c.
        \end{equation}
        
        We argue that the probability that the variance $\sigma_2^2$ of ${\bm{w}}_2$ is unequal to the variance $\sigma_1^2$ of ${\bm{w}}_1$ is $1$. 
        That is, almost surely, 
        \begin{equation}
            \label{eq:variance_equal}
            \mathop {\lim }\limits_{N \to \infty } \frac{1}{N} \sum\limits_{i = 1}^N  w_{1, i}^2 \ne \mathop {\lim }\limits_{N \to \infty } \frac{1}{N} \sum\limits_{i = 1}^N w_{2, i}^2 - \mu_2^2, 
        \end{equation}
        where
        \begin{equation}
            \mu_2 = \frac{1}{N} \sum\limits_{i = 1}^N w_{2, i}.
        \end{equation}
        Such conclusion holds since the well-known fact that the set of zeros of a nonzero polynomial has measure zero. 
        
        Without loss of generality, let $\sigma_2 > \sigma_1$.
        Denote the empirical distribution of ${\bm{w}}_1(N)$ and ${\bm{w}}_2(N)$ by $F_{1, N}(x)$ and $F_{2, N}(x)$ respectively, given by
        \begin{eqnarray}
            \label{eq:F_1N}
            &&F_{1, N}(x) = \frac{1}{N} \# \left\{ i \le N : w_{1, i} \le x \right\}, \\
            \label{eq:F_2N}
            &&F_{2, N}(x) = \frac{1}{N} \# \left\{ i \le N : w_{2, i} \le x \right\}.
        \end{eqnarray}
        Therefore, $\forall \delta_1 > 0$, $\forall \delta_2 > 0$, $\forall m \in {\mathbb{R}}$, $\exists N_1 \in {\mathbb{N}}$, $\forall n \ge N_1$, such that
        \begin{eqnarray}
            \label{eq:F_1N_F_1}
            &&\left| F_{1, n}(m \sigma_1) - F_{1}(m \sigma_1) \right| \le \delta_1, \\
            \label{eq:F_2N_F_2}
            &&\left| F_{2, n}(m \sigma_2 + \mu_2) - F_{2}(m \sigma_2 + \mu_2) \right| \le \delta_2
        \end{eqnarray}
        hold with probability $1$, where 
        \begin{equation}
            F_{1}(m \sigma_1) = F_2(m \sigma_2 + \mu_2) = \frac{1}{\sqrt{2 \pi}} {\int_{-\infty}^m {\rm{e}}^{-\frac{t^2}{2}} {\rm{d}}t }.
        \end{equation}
        Thus, 
        \begin{equation}
            \left| F_{1, n}(m \sigma_1) - F_{2, n}(m \sigma_2 + \mu_2) \right| \le \delta_1 + \delta_2.
        \end{equation}
        Let $m > \frac{c - \mu_2}{\sigma_2 - \sigma_1}$, $m_1 = \frac{m \sigma_2 + \mu_2 - c}{\sigma_1}$, and
        \begin{equation}
            \delta_1 = \delta_2 = \frac{1}{\sqrt{18 \pi}} (m_1 - m) {\rm{e}}^{-\frac{m_1^2}{2}}.
        \end{equation}
        Then we have $m \sigma_2 + \mu_2 = m_1 \sigma_1 + c$ and $m_1 > m$.
        Combining with \eqref{eq:w_2i_w_1i} and the definition of the empirical law \eqref{eq:F_1N}, \eqref{eq:F_2N} yields 
        \begin{equation}
            \label{eq:F2geF1}
            F_{2, n}(m \sigma_2 + \mu_2) \ge F_{1, n}(m_1 \sigma_1).
        \end{equation}
        On the other hand, \eqref{eq:F_1N_F_1} and \eqref{eq:F_2N_F_2} yield
        \begin{eqnarray}
            &&F_{1, n}(m_1 \sigma_1) \ge \frac{1}{\sqrt{2 \pi}} {\int_{-\infty}^{m_1} {\rm{e}}^{-\frac{t^2}{2}} {\rm{d}}t } - \delta_1, \\
            &&F_{2, n}(m \sigma_2 + \mu_2) \le \frac{1}{\sqrt{2 \pi}} {\int_{-\infty}^{m} {\rm{e}}^{-\frac{t^2}{2}} {\rm{d}}t } + \delta_2.
        \end{eqnarray}
        Then, from a simple math, we can obtain that 
        \begin{equation}
            F_{2, n}(m \sigma_2 + \mu_2) - F_{1, n}(m_1 \sigma_1) \le \delta_1 + \delta_2 - \frac{1}{\sqrt{2 \pi}} {\int_{m}^{m_1} {\rm{e}}^{-\frac{t^2}{2}} {\rm{d}}t } < 0, 
        \end{equation}
        which is in contradiction with \eqref{eq:F2geF1}. 
    
    \section{Proof of Theorem \ref{theorem:detection_performance}}
    \label{appendix:performance}
    
        For the convenience of readers, we here restate the definition of subdifferential as follow, which will help us prove this theorem.
        
        \begin{definition}[\cite{murphy2012machine}]
	        
	        \label{def:subdifferential}
	        
	        The subdifferential of a function $f : {\cal I} \to {\mathbb{R}}$ at a point ${\bm x}_0 \in {\cal I}$ is defined to be a set of vectors such that
            \begin{equation}
                {\left. {\partial f \left( {\bm x} \right)} \right|_{{\bm x} = {{\bm x}_0}}} = \left\{ {\bm g} \in {\mathbb{R}}^n : f \left( {\bm x} \right) - f \left( {\bm x}_0 \right) \ge \left( {\bm x} - {\bm x}_0 \right)^T {\bm g}, \quad \forall {\bm x} \in {\cal I} \right\},
            \end{equation}
            where ${\cal I} \subset {\mathbb{R}}^n$ and each ${\bm g}$ is called subderivative or subgradient.
	        
	    \end{definition}
	    
	    It is easy to verify that the subderivative is linear.
	    We then prove the following lemma that simplifies the proof of Theorem \ref{theorem:detection_performance}.
	    
	    \begin{lemma}
	    
	        \label{lemma:solution_condition}
	        
	        The vector $\hat {\bm{x}}$ is the solution of the \ac{lasso} problem if and only if the following conditions are all satisfied for $i \in \left\{ 1, \ldots, N \right\}$: 
	        \begin{eqnarray}
	            \label{eq:condition_ne0}
	            &&{\bm{a}}_i^H \left( {\bm{y}} - {\bm{A}} {\hat {\bm{x}}} \right) = \lambda \frac{\hat x_i}{\left| \hat x_i \right|}, \quad {\text{if  }} \hat x_i \ne 0,\\
	            \label{eq:condition_eq0}
	            &&\left| {\bm{a}}_i^H \left( {\bm{y}} - {\bm{A}} {\hat {\bm{x}}} \right) \right| \le \lambda, \quad {\text{if  }} \hat x_i = 0,
	        \end{eqnarray}
	        where ${\bm{a}}_i$ is the $i$-th column of matrix ${\bm{A}}$ and $\hat x_i$ is the $i$-th entry of vector $\hat {\bm{x}}$. 
	        
	        \begin{proof}
	            
	            Due to the independent variable ${\bm{x}}$ in \ac{lasso} problem is complex-valued, we transform the objective function into a real function $f : {\mathbb{R}}^{2N} \to {\mathbb{R}}$, given by
	            \begin{equation}
	                f \left( {\bm{x}}_r \right) = \frac{1}{2} {\left\| {\bm y}_r - {\bm A}_r {\bm x}_r \right\|_2^2} + \lambda {\sum\limits_{i = 1}^N \left( x_{r, i}^2 + x_{r, i + N}^2 \right)^{\frac{1}{2}}},
	            \end{equation}
	            where ${\bm x}_r = [ {\rm{Re}} \left( {\bm x} \right)^T, {\rm{Im}} \left( {\bm x} \right)^T ]^T \in {\mathbb{R}}^{2N}$, ${\bm y}_r = [ {\rm{Re}} \left( {\bm y} \right)^T, {\rm{Im}} \left( {\bm y} \right)^T ]^T \in {\mathbb{R}}^{2M}$ and 
	            \[ 
	            {\bm A}_r = \left[ 
	            {\begin{array}{*{20}{l}}
                {\rm{Re}} \left( {\bm A} \right) & -{\rm{Im}} \left( {\bm A} \right)\\
                {\rm{Im}} \left( {\bm A} \right) & {\rm{Re}} \left( {\bm A} \right)
                \end{array}}
                \right] \in {\mathbb{R}}^{2M \times 2N}.
                \]
                It is easy to verify that $f \left( {\bm{x}}_r \right)$ equals to the objective function of \ac{lasso} problem. 
                Let $f \left( {\bm{x}}_r \right) = g \left( {\bm{x}}_r \right) + \lambda h \left( {\bm{x}}_r \right)$, in which
                \begin{eqnarray}
                    &&g \left( {\bm{x}}_r \right) = \frac{1}{2} {\left\| {\bm y}_r - {\bm A}_r {\bm x}_r \right\|_2^2}, \\
                    &&h \left( {\bm{x}}_r \right) = {\sum\limits_{i = 1}^N \left( x_{r, i}^2 + x_{r, i + N}^2 \right)^{\frac{1}{2}}}.
                \end{eqnarray}
                The part $g \left( {\bm{x}}_r \right)$ is everywhere differentiable, thus 
                \begin{equation}
                    \partial g \left( {\bm{x}}_r \right) = \left\{ \frac{{\rm{d}} g \left( {\bm{x}}_r \right)}{{\rm{d}} {\bm{x}}_r } \right\} = \left\{ - {\bm A}_r \left( {\bm y}_r - {\bm A}_r {\bm x}_r \right) \right\}.
                \end{equation}
                The part $h \left( {\bm{x}}_r \right)$ could be divided into $N$ parts: $h_i \left( x_{r, i}, x_{r, i + N} \right) = ( x_{r, i}^2 + x_{r, i + N}^2 )^{\frac{1}{2}}$. 
                If $x_{r, i}^2 + x_{r, i + N}^2 > 0$, then $\partial h_i \left( x_{r, i}, x_{r, i + N} \right) = \{ ( x_{r, i} ( x_{r, i}^2 + x_{r, i + N}^2 )^{-\frac{1}{2}}, x_{r, i + N} ( x_{r, i}^2 + x_{r, i + N}^2 )^{-\frac{1}{2}} ) \}$. 
                If $x_{r, i}^2 + x_{r, i + N}^2 = 0$, then $\partial h_i \left( x_{r, i}, x_{r, i + N} \right) = \{ (g_1, g_2) : ( x_{r, i}^2 + x_{r, i + N}^2 )^{\frac{1}{2}} \ge g_1 x_{r, i} + g_2 x_{r, i + N} \}$.
                According to Cauchy inequality, we can easily obtain that $g_1^2 + g_2^2 \le 1$.
                
                With the fact that $f \left( {\bm{x}}_r \right)$ is a convex function, the point $\hat {\bm{x}}_r$ is the global minimum of $f \left( {\bm{x}}_r \right)$ if and only if ${\bm 0} \in {\left. {\partial f \left( {\bm x}_r \right)} \right|_{{\bm x}_r = \hat {\bm{x}}_r}}$, which indicates that
                \begin{equation}
                    {\bm l} \buildrel \textstyle. \over = \frac{1}{\lambda} {\bm A}_r \left( {\bm y}_r - {\bm A}_r {\bm x}_r \right) \in {\left. {\partial h \left( {\bm x}_r \right)} \right|_{{\bm x}_r = \hat {\bm{x}}_r}}
                \end{equation}
                Therefore, 
                \begin{equation}
                    \left( l_i, l_{i + N} \right)  \in \left\{ {\begin{array}{*{20}{l}}
                    {\left\{ \left( \frac{x_{r, i}}{( x_{r, i}^2 + x_{r, i + N}^2 )^{\frac{1}{2}}}, \frac{x_{r, i + N}}{( x_{r, i}^2 + x_{r, i + N}^2 )^{\frac{1}{2}}} \right) \right\},} & {x_{r,i}^2 + x_{r,i + N}^2 > 0;}\\
                    {\left\{ {\left( {{g_1},{g_2}} \right) : g_1^2 + g_2^2 \le 1} \right\},} & {x_{r,i}^2 + x_{r,i + N}^2 = 0.}
                    \end{array}} \right.
                \end{equation}
                Combined with the definition of ${\bm x}_r$, ${\bm y}_r$ and ${\bm A}_r$, the previous result yields \eqref{eq:condition_ne0} and \eqref{eq:condition_eq0}.
	            
	        \end{proof}
	        
	    \end{lemma}
        
        Let the threshold of the \ac{lasso} detector be $\kappa$, such that the probability of false alarm $P_{fa, 2}$ satisfies
        \begin{equation}
             P_{fa, 2} = \mathop {\lim }\limits_{N \to \infty } \frac{1}{N-k} \sum\limits_{i \in S^c} { \varphi_{2, i} },
        \end{equation}
        where $S = \left\{ i: x_{0, i} \ne 0 \right\}$ is the support set of ${\bm x}_0$ with $|S| = k$, $S^c = \{1, \ldots, N\} \backslash S$ and $\varphi_{2, i}$ is a test with
        \begin{equation}
	        {\varphi_{2, i}} = \left\{ {\begin{array}{*{20}{l}}
            {1,}&{\left| \hat {x}^{\rm{LASSO}}_i \right| > \kappa;}\\
            {0,}&{{\text{otherwise.}}}
            \end{array}} \right.
	    \end{equation}
	    Then the probability of detection of \ac{lasso} detector $P_{d, 2}$ has the expression that
	    \begin{equation}
	        P_{d, 2} = \mathop {\lim }\limits_{N \to \infty } \frac{1}{k} \sum\limits_{i \in S} { \varphi_{2, i} }.
	    \end{equation}
	    We now define the test of the debiased \ac{lasso} detector as follow: 
	    \begin{equation}
	        {\varphi_{1, i}} = \left\{ {\begin{array}{*{20}{l}}
            {1,}&{\left| \hat {x}^{\rm{d}}_i \right| > \kappa + \frac{\lambda}{\Lambda};}\\
            {0,}&{{\text{otherwise.}}}
            \end{array}} \right.
	    \end{equation}
	    and the probability of false alarm $P_{fa, 1}$, the probability of detection $P_{d, 1}$, given by
	    \begin{eqnarray}
	        &&P_{fa, 1} = \mathop {\lim }\limits_{N \to \infty } \frac{1}{N-k} \sum\limits_{i \in S^c} { \varphi_{1, i} },\\
	        &&P_{d, 1} = \mathop {\lim }\limits_{N \to \infty } \frac{1}{k} \sum\limits_{i \in S} { \varphi_{1, i} }.
	    \end{eqnarray}
	    
	    \begin{enumerate}[1)]
	    
	        \item 
	        If $\varphi_{2, i} = 0$ and $i \in S^c$, then $\left| \hat {x}^{\rm{LASSO}}_i \right| \le \kappa$.
	        \begin{itemize}
	            
	            \item
    	        For $i$ satisfying $\hat{x}^{\rm{LASSO}}_i = 0$, we have the following derivation based on Lemma \ref{lemma:solution_condition}: 
    	        \begin{eqnarray}
    	            \left| \hat {x}^{\rm{d}}_i \right| &=& \left| \hat {x}^{\rm{LASSO}}_i + \frac{1}{\Lambda} {\bm{a}}_i^H \left( {\bm{y}} - {\bm{A}} {\hat {\bm{x}}^{\rm{LASSO}}} \right) \right| \nonumber \\
    	            &\le& \left| \hat {x}^{\rm{LASSO}}_i \right| +  \frac{1}{\Lambda} \left| {\bm{a}}_i^H \left( {\bm{y}} - {\bm{A}} {\hat {\bm{x}}^{\rm{LASSO}}} \right) \right| \nonumber \\
    	            &\le& \left| \hat {x}^{\rm{LASSO}}_i \right| + \frac{\lambda}{\Lambda}.
    	        \end{eqnarray}
    	        
    	        \item
    	        For $i$ satisfying $\left| \hat{x}^{\rm{LASSO}}_i \right| > 0$, we have the following derivation based on Lemma \ref{lemma:solution_condition}: 
    	        \begin{eqnarray}
    	            \left| \hat {x}^{\rm{d}}_i \right| &=& \left| \hat {x}^{\rm{LASSO}}_i + \frac{1}{\Lambda} {\bm{a}}_i^H \left( {\bm{y}} - {\bm{A}} {\hat {\bm{x}}^{\rm{LASSO}}} \right) \right| \nonumber \\
    	            &=& \left| \hat {x}^{\rm{LASSO}}_i  +  \frac{\lambda}{\Lambda} \cdot \frac{\hat x^{\rm{LASSO}}_i}{\left| \hat x^{\rm{LASSO}}_i \right|}  \right| \nonumber \\
    	            &=& \left| \hat {x}^{\rm{LASSO}}_i \right| + \frac{\lambda}{\Lambda}.
    	        \end{eqnarray}
    	        
	        \end{itemize}
	        
	        Therefore, $\left| \hat {x}^{\rm{d}}_i \right| \le \kappa + \frac{\lambda}{\Lambda}$ and $\varphi_{1, i} = 0$.
	        Thus, $\varphi_{1, i} \le \varphi_{2, i}$ for $i \in S^c$, and
	        \begin{eqnarray}
	            P_{fa, 1} &=& \mathop {\lim }\limits_{N \to \infty } \frac{1}{N-k} \sum\limits_{i \in S^c} { \varphi_{1, i} } \nonumber \\
	            &\le& \mathop {\lim }\limits_{N \to \infty } \frac{1}{N-k} \sum\limits_{i \in S^c} { \varphi_{2, i} } = P_{fa, 2}.
	        \end{eqnarray}
	        
	        \item
	        If $\varphi_{2, i} = 1$ and $i \in S$, then $\left| \hat {x}^{\rm{LASSO}}_i \right| > \kappa$, leading to $\left| \hat {x}^{\rm{d}}_i \right| > \kappa + \frac{\lambda}{\Lambda}$ and $\varphi_{1, i} = 1$.
	        Thus, $\varphi_{1, i} \ge \varphi_{2, i}$ for $i \in S$, and
	        \begin{eqnarray}
	            P_{d, 1} &=& \mathop {\lim }\limits_{N \to \infty } \frac{1}{k} \sum\limits_{i \in S} { \varphi_{1, i} } \nonumber \\
	            &\ge& \mathop {\lim }\limits_{N \to \infty } \frac{1}{k} \sum\limits_{i \in S} { \varphi_{2, i} } = P_{d, 2}.
	        \end{eqnarray}

	    \end{enumerate}

	\section{Proof of Claim \ref{clm:FED}}
	\label{appendix:derivation}
	    
	    Replica method bases on a simple fact that for any $Z > 0$, we have
	    \begin{equation}
	        \label{eq:rm1}
            \ln Z = \mathop {\lim }\limits_{n \to 0} \frac{{{Z^n} - 1}}{n} = \mathop {\lim }\limits_{n \to 0} \frac{\partial }{{\partial n}}\left( {{Z^n} - 1} \right).
        \end{equation}
        If $Z$ is a random variable, then we take the average over $Z$ on the both side of \eqref{eq:rm1} yielding
        \begin{equation}
	        \label{eq:rm2}
            \mathbb{E} \ln Z = \mathop {\lim }\limits_{n \to 0} \frac{{{\mathbb{E} Z^n} - 1}}{n} = \mathop {\lim }\limits_{n \to 0} \frac{1}{n}\ln \left( {{\mathbb{E} Z^n}} \right) = \mathop {\lim }\limits_{n \to 0} \frac{\partial }{{\partial n}}\ln \left( {{\mathbb{E} Z^n}} \right).
        \end{equation}
        Therefore, with the help of replica method, we rewrite the free energy density $f(\lambda)$ as follow
    	\begin{equation}
    	    f(\lambda) = - \mathop {\lim }\limits_{\beta  \to \infty } \mathop {\lim }\limits_{N \to \infty } \mathop {\lim }\limits_{n \to 0 } \frac{\partial }{{\partial n}} \frac{1}{ \beta N} \ln {\mathbb{E}}_{{\bm A}, {\bm \xi}} \left[ Z^n \left( {{\bm y}, {\bm A}; \lambda, \beta} \right) \right],
    	\end{equation}
    	where we recall $Z \left( {{\bm y}, {\bm A}; \lambda, \beta} \right)$ is the partition function and we use $Z$ instead of $Z \left( {{\bm y}, {\bm A}; \lambda, \beta} \right)$ in all of the following.
    	As most of the work based on replica analysis, we exchange the order of the limits $n \to 0$ and $N \to \infty$, thus
    	\begin{equation}
    	    f(\lambda) = - \mathop {\lim }\limits_{\beta  \to \infty } \mathop {\lim } \limits_{n \to 0 } \frac{\partial }{{\partial n}} \mathop {\lim }\limits_{N \to \infty }  \frac{1}{ \beta N} \ln {\mathbb{E}}_{{\bm A}, {\bm \xi}} \left[ Z^n \right].
    	\end{equation}
    	We will first compute the limit $N \to \infty$ in Appendix \ref{subsec:N}, then we take the other limits to get the final result in Appendix \ref{subsec:final}.
    	
	    \subsection{Derivation of the limit $N \to \infty$}
	    \label{subsec:N}
    	
    	In this subsection, we will introduce the derivation of 
    	\[\frac{1}{N} \ln {\mathbb{E}}_{{\bm A}, {\bm \xi}} \left[ Z^n \right]\]
    	and take the limit $N \to \infty$. 
    	With the definition of $Z$, we have
    	\small 
        \begin{eqnarray}
    	    \label{eq:EZn}
    	    &&\frac{1}{N} \ln {\mathbb{E}}_{{\bm A}, {\bm \xi}} \left[ Z^n \right] \nonumber \\ 
    	    &=& \frac{1}{N} \ln {\mathbb{E}}_{{\bm A}, {\bm \xi}} \left[ \int{ {\prod \limits_{a = 1}^n   {\rm d}{\bm x}_a \exp \left\{ -\frac{\beta}{2} \left\| {\bm y} - {\bm A}{\bm x}_a \right\|_2^2 - \beta \lambda {\left\| {\bm x}_a \right\|_1} \right\} } } \right] \nonumber \\ 
    	    &=& \frac{1}{N} \ln \int{ {\prod \limits_{a = 1}^n \exp \left\{ -\frac{\beta}{2} \left\| {\bm y} - {\bm A}{\bm x}_a \right\|_2^2 - \beta \lambda {\left\| {\bm x}_a \right\|_1} \right\} } p_{\bm \xi}({\bm \xi}) p_{\bm A}({\bm A}) {\prod \limits_{a = 1}^n   {\rm d}{\bm x}_a} {\rm d}{\bm \xi} {\rm d}{\bm A}},
    	\end{eqnarray}
    	\normalsize
    	where $p_{\bm{A}} ({\bm{A}})$ and $p_{\bm \xi} ({\bm \xi})$ are the probability density function of ${\bm{A}}$ and ${\bm \xi}$, respectively.
    	However, this integral is still difficult to calculate.
    	To solve this problem, we introduce the \ac{rs} ansatz, which gives a constraint on original signal ${\bm{x}}_0$ and the replicas ${\bm{x}}_a$, and we express it as
    	\begin{equation}
    	\label{eq:RS_h}
    	    {\bm{h}}\left( \left\{ {{{\bm{x}}_a}} \right\}_{a = 1}^n,{{\bm{x}}_0} \right) = {\bm{k}}.
    	\end{equation}
        And we define that
        \begin{equation}
            f\left( \left\{ {{{\bm{x}}_a}} \right\}_{a = 1}^n, {{\bm{x}}_0}, {\bm A}, {\bm \xi} \right) \buildrel \textstyle. \over = {\prod \limits_{a = 1}^n \exp \left\{ -\frac{\beta}{2} \left\| {\bm y} - {\bm A}{\bm x}_a \right\|_2^2 \right\}} p_{\bm \xi}({\bm \xi}) p_{\bm A}({\bm A}),
        \end{equation}
        \begin{equation}
            g\left( \left\{ {{{\bm{x}}_a}} \right\}_{a = 1}^n \right) \buildrel \textstyle. \over = {\prod \limits_{a = 1}^n{\rm e}^{- \beta \lambda {\left\| {\bm x}_a \right\|_1}}}.
        \end{equation}
        With the conclusion
        \begin{equation}
    	    \label{eq:intf}
    	    \int{ f\left( \left\{ {{{\bm{x}}_a}} \right\}_{a = 1}^n, {{\bm{x}}_0}, {\bm A}, {\bm \xi} \right) {\rm d}{\bm \xi} {\rm d}{\bm A} } = f_1 \left( {\bm{h}}\left( \left\{ {{{\bm{x}}_a}} \right\}_{a = 1}^n,{{\bm{x}}_0} \right) \right),
    	\end{equation}
    	which will be proved in Appendix \ref{subsubsec:Tk}, we can split \eqref{eq:EZn} into two parts under the \ac{rs} ansatz:
        \small 
        \begin{eqnarray}
            \label{eq:fg}
            &&\frac{1}{N} \ln {\mathbb{E}}_{{\bm A}, {\bm \xi}} \left[ Z^n \right] \nonumber \\  
            &=& \frac{1}{N} \ln \int{ f\left( \left\{ {{{\bm{x}}_a}} \right\}_{a = 1}^n, {{\bm{x}}_0}, {\bm A}, {\bm \xi} \right) g\left( \left\{ {{{\bm{x}}_a}} \right\}_{a = 1}^n \right) \delta \left( {\bm{h}}\left( \left\{ {{{\bm{x}}_a}} \right\}_{a = 1}^n,{{\bm{x}}_0} \right) - {\bm{k}} \right) {\prod \limits_{a = 1}^n   {\rm d}{\bm x}_a}} {\rm d}{\bm \xi} {\rm d}{\bm A}   {\rm d}{\bm k} \nonumber \\
            &=& \frac{1}{N} \ln \int{ f_1\left( {\bm{h}}\left( \left\{ {{{\bm{x}}_a}} \right\}_{a = 1}^n,{{\bm{x}}_0} \right) \right) g\left( \left\{ {{{\bm{x}}_a}} \right\}_{a = 1}^n \right) \delta \left( {\bm{h}}\left( \left\{ {{{\bm{x}}_a}} \right\}_{a = 1}^n,{{\bm{x}}_0} \right) - {\bm{k}} \right) {\prod \limits_{a = 1}^n   {\rm d}{\bm x}_a}  } {\rm d}{\bm k} \nonumber \\
            &=& \frac{1}{N} \ln \int f_1\left( {\bm{k}} \right) \left( \int{  g\left( \left\{ {{{\bm{x}}_a}} \right\}_{a = 1}^n \right) \delta \left( {\bm{h}}\left( \left\{ {{{\bm{x}}_a}} \right\}_{a = 1}^n,{{\bm{x}}_0} \right) - {\bm{k}} \right) {\prod \limits_{a = 1}^n   {\rm d}{\bm x}_a} } \right) {\rm d}{\bm k}  \nonumber \\
    	    &=& \frac{1}{N} \ln \int f_1\left( {\bm{k}} \right) g_1\left( {\bm{k}}, {{\bm{x}}_0} \right) {\rm d}{\bm k},
    	\end{eqnarray} 
    	\normalsize
    	where 
    	\begin{equation}
    	    \label{eq:intg}
    	    g_1\left( {\bm{k}}, {{\bm{x}}_0} \right) \buildrel \textstyle. \over = \int{ g\left( \left\{ {{{\bm{x}}_a}} \right\}_{a = 1}^n \right) \delta \left( {\bm{h}}\left( \left\{ {{{\bm{x}}_a}} \right\}_{a = 1}^n, {{\bm{x}}_0} \right) - {\bm{k}} \right) {\prod \limits_{a = 1}^n {\rm d}{\bm x}_a} }.
    	\end{equation}
    	Then we employ the following substitution, given by
    	\begin{eqnarray}
    	    f_1\left( {\bm{k}} \right) = \exp \left\{ N {\cal T} \left( {\bm{k}} \right) \right\}, \\
    	    g_1\left( {\bm{k}}, {\bm x}_0 \right) = \exp \left\{ N {\cal S} \left( {\bm{k}}, {\bm x}_0 \right) \right\}.
    	\end{eqnarray}
    	The limit $N \to \infty$ of \eqref{eq:fg} can be calculated by saddle point method, which is usually used to approximately calculate the following complex integral when $N \in {\mathbb{R}}$ is large enough
        \begin{equation}
            I(N) = {\int_{\Gamma_1 \times \Gamma_2 \times \ldots \times \Gamma_M} F({\bm z}) {\rm e}^{N f({\bm z})} {\rm{d}} {\bm z} },
        \end{equation}
        where $f({\bm z})$ is analytic in complex variable ${\bm z} \in {\mathbb{C}}^{M}$ and $\left\{ \Gamma_i \right\}$ converge the integral.
        Saddle point method claims that for large $N$, the integral can be approximately given by the function value of the saddle point ${\bm z}_0$,
        given by
        \begin{equation}
            \left| I(N) \right| = \left| {\rm e}^{N \left( f({\bm z}_0) + O \left( \frac{1}{N} \right) \right) } \right| ,
        \end{equation}
        where ${\bm z}_0$, the saddle point, satisfies $\left. \frac{\partial f}{\partial {\bm z}} \right|_{{\bm z} = {\bm z}_0} = {\bm 0}$.
        Thus, 
        \begin{equation}
            \label{eq:T+S}
            \mathop {\lim }\limits_{N \to \infty } \frac{1}{N} \ln {\mathbb{E}}_{{\bm A}, {\bm \xi}} \left[ Z^n \right] = \mathop {\rm {extr}}\limits_{\bm k} \left\{ {\cal T} \left( {\bm{k}} \right) + {\cal S} \left( {\bm{k}}, {\bm x}_0 \right) \right\}.
        \end{equation}
    	In Appendix \ref{subsubsec:Tk} and \ref{subsubsec:Sk}, we will provide the derivation of ${\cal T} \left( {\bm{k}} \right)$ and ${\cal S} \left( {\bm{k}}, {\bm x}_0 \right)$, respectively.
    	
    	\subsubsection{The derivation of ${\cal T}({\bm k})$}
    	\label{subsubsec:Tk}
        	
        	In this section, we will introduce the derivation of
        	\begin{equation}
        	    \label{eq:intf2}
        	    \int{ f\left( \left\{ {{{\bm{x}}_a}} \right\}_{a = 1}^n, {{\bm{x}}_0}, {\bm A}, {\bm \xi} \right) {\rm d}{\bm \xi} {\rm d}{\bm A} } = {\mathbb{E}}_{{\bm A}, {\bm \xi}} \left[ \exp \left\{ -\frac{\beta}{2} {\sum \limits_{a = 1}^n \left\| {\bm y} - {\bm A}{\bm x}_a \right\|_2^2 } \right\}  \right],
        	\end{equation}
        	and further under the \ac{rs} ansatz
        	\begin{equation}
        	    {\cal T}({\bm k}) = \mathop {\lim }\limits_{N \to \infty} \frac{1}{N} \ln f_1\left( {\bm{k}} \right) = \mathop {\lim }\limits_{N \to \infty} \frac{1}{N} \ln \int{ f\left( \left\{ {{{\bm{x}}_a}} \right\}_{a = 1}^n, {{\bm{x}}_0}, {\bm A}, {\bm \xi} \right) {\rm d}{\bm \xi} {\rm d}{\bm A} }.
        	\end{equation}
        	
        	It is convenient to first take the average with respect to $\bm \xi$: 
        	\begin{equation}
        	    \label{eq:Exi}
        	    {\mathbb{E}}_{{\bm \xi}} \left[ \exp \left\{ -\frac{\beta}{2} {\sum \limits_{a = 1}^n \left\| {\bm A} \left( {\bm x}_0 - {\bm x}_a \right) + {\bm \xi} \right\|_2^2 } \right\} \right] = \exp \left[ {\sum \limits_{a = 0}^n -\frac{k_a}{2} {\bm u}_a^H {\bm{J}}{\bm u}_a  } - N \gamma \ln \left( 1 + \frac{\beta n \sigma^2}{2} \right) \right],
        	\end{equation}
        	where we recall that ${\bm{J}} = {\bm{A}}^H {\bm{A}}$, $\gamma = M / N$ and $k_a$, ${\bm{u}}_a$ are defined as follows: 
            \begin{eqnarray}
                \label{eq:ku}
                k_0 &=& -\frac{\beta^2 \sigma^2}{2 + \beta n \sigma^2}, \\
                {\bm u}_0 &=& {\sum \limits_{a = 1}^n {\bm u}_a}, \\
                k_a &=& \beta, \quad 1 \le a \le n, \\
                \label{eq:u_a}
                {\bm u}_a &=& {\bm x}_a - {\bm x}_0,  \quad 1 \le a \le n.
            \end{eqnarray}
            Then we orthogonalize $\{ {\bm u}_a \}_{a=0}^n$ in order to facilitate the calculation of the average on $\bm A$.
            First, one can rewrite \eqref{eq:Exi} as,
            \begin{equation}
                \exp \left[ {\sum \limits_{a = 0}^n -\frac{k_a}{2} {\bm u}_a^H {\bm{J}}{\bm u}_a  } - N \gamma \ln \left( 1 + \frac{\beta n \sigma^2}{2} \right) \right] = \exp \left[ \frac{1}{2} {\rm{Tr}} \left( {\bm{J}} {\bm{L}} \right) - N \gamma \ln \left( 1 + \frac{\beta n \sigma^2}{2} \right) \right],
            \end{equation}
            where ${\rm{Tr}}({\cdot})$ denotes the trace of a matrix and ${\bm{L}} \in \mathbb{C}^{N \times N}$ is defined as follow: 
            \begin{equation}
                \label{eq:L}
                {\bm{L}} = \frac{\beta^2 \sigma^2}{2 + \beta n \sigma^2} \left( {\sum \limits_{a = 1}^n {\bm u}_a } \right) \left( {\sum \limits_{a = 1}^n {\bm u}_a } \right)^H - \beta {\sum \limits_{a = 1}^n {\bm u}_a {\bm u}_a^H }.
            \end{equation}
            To obtain a more specific perspective on the matrix $\bm L$, we here introduce the \ac{rs} ansatz.
            Denote by an overlap matrix ${\bm Q} \in {\mathbb C}^{(n+1) \times (n+1)}$, whose elements are given by $Q_{ab} = Q_{ba} = (2N)^{-1}  {\bm{x}}_a^H {\bm{x}}_b$ $(0 \le a \le n; 0 \le b \le n)$.
            Then the \ac{rs} ansatz restricts the values of the overlap matrix ${\bm Q}$ to the following: 
            \begin{equation}
            \label{eq:Qab}
                {Q_{ab}} = \left\{ {\begin{array}{*{20}{l}}
                {\begin{array}{*{20}{l}}
                {Q,}\\
                {q,}\\
                {m,}\\
                {\rho ,}
                \end{array}}&{\begin{array}{*{20}{l}}
                {(a = b = 1,2, \ldots ,n)}\\
                {(a > b = 1,2, \ldots ,n;b > a = 1,2, \ldots ,n)}\\
                {(a = 0,b = 1,2, \ldots ,n)}\\
                {(a = b = 0)}
                \end{array}}
                \end{array}}, \right.
            \end{equation}
            where we recall $\rho > 0$ is half of the signal density,
            $Q$, $m$ and $q$ are real numbers.
            Due to the symmetry of the replicas, we assume that ${\bm{x}}_a^H {\bm{x}}_b$ for $a \ne b$ are all real numbers.
            This indicates that ${\bm L}$ has three types of eigenvalues: 
            \begin{eqnarray}
                s_1 &=& -\frac{4N \beta}{2 + \beta n \sigma^2} \left( (Q - q) + n (q - 2m + \rho) \right),\\
                s_2 &=& - 2N \beta (Q - q),\\
                s_3 &=& 0,
            \end{eqnarray}
            which can be obtained by
            \begin{eqnarray}
                {\bm L} {\sum \limits_{a = 1}^n {\bm u}_a } &=& -\frac{4N \beta}{2 + \beta n \sigma^2} \left( (Q - q) + n (q - 2m + \rho) \right) {\sum \limits_{a = 1}^n {\bm u}_a },\\
                {\bm L} \left( {\sum \limits_{a = 1}^n {\bm u}_a } - n {\bm u}_b \right) &=& - 2N \beta (Q - q) \left( {\sum \limits_{a = 1}^n {\bm u}_a } - n {\bm u}_b \right),\\
                {\bm L} {\bm v}_k &=& {\bm 0},
            \end{eqnarray}
            where ${\bm v}_k$ are the vector orthogonal to all ${\bm u}_a$, thus there are at least $N-n$ ${\bm v}_k$'s.
            Now we get the eigenvalue decomposition of $\bm L$, thus  
            \begin{equation}
        	    \label{eq:uJu_vJv}
        	     \exp \left[ {\sum \limits_{a = 0}^n -\frac{k_a}{2} {\bm u}_a^H {\bm{J}}{\bm u}_a  } \right] = \exp \left[ {\sum \limits_{i = 1}^n -\frac{1}{2} {\bm v}_i^H {\bm{J}}{\bm v}_i} \right],
        	\end{equation}
        	where 
        	\begin{eqnarray}
                {\bm v}_1^H {\bm v}_1 &=& 2N \cdot \frac{2\beta}{2 + \beta n \sigma^2} \left( (Q - q) + n (q - 2m + \rho) \right),\\
                {\bm v}_i^H {\bm v}_i &=&  2N \cdot \beta (Q - q), \quad 2 \le i \le n.
            \end{eqnarray}
            
            Under such constraint, when $N$ is large enough, the logarithm of the averages on ${\bm A}$ can be calculated:
            \begin{eqnarray}
                \label{eq: TQqm}
                &&{\cal T}({\bm k}) \nonumber \\
                &\buildrel \textstyle. \over =&{\cal T}(Q, q, m) \nonumber \\
                &=&\mathop {\lim }\limits_{N \to \infty} \frac{1}{N} \ln \int{ f\left( \left\{ {{{\bm{x}}_a}} \right\}_{a = 1}^n, {{\bm{x}}_0}, {\bm A}, {\bm \xi} \right) {\rm d}{\bm \xi} {\rm d}{\bm A} } \nonumber \\
                &=& \mathop {\lim }\limits_{N \to \infty} \frac{1}{N} \ln {\mathbb{E}}_{\bm A} \left\{ \exp \left[ {\sum \limits_{i = 1}^n -\frac{1}{2} {\bm v}_i^H {\bm{J}}{\bm v}_i} - \frac{N \gamma}{2} \ln \left( 1 + \beta n \sigma^2 \right) \right\} \right] \nonumber \\
                &=& G \left( - \frac{2\beta \left( Q - q + n(q - 2m + \rho) \right)}{2 + \beta n \sigma^2}; {\bm J} \right) \nonumber \\
                && + (n-1) G(-\beta(Q-q); {\bm J}) - \gamma \ln \left( 1 + \frac{\beta n \sigma^2}{2} \right),
            \end{eqnarray}
            where
            \begin{equation}
                G(x; {\bm{J}}) \buildrel \textstyle. \over = \mathop {\rm{extr}} \limits_{z} \left[ -\int{\rho_{\bm J}(s) \ln \left| z - s \right| {\rm{d}}s} + zx \right] - \ln \left| x \right| - 1.
            \end{equation}
            The derivation of \eqref{eq: TQqm} is explained in Appendix \ref{appendix:ROM}.
    	
    	\subsubsection{The derivation of ${\cal S} ({\bm k}, {\bm x}_0)$}
    	\label{subsubsec:Sk}
    	    
    	    In this section, we will derive 
    	    \begin{equation}
    	        {\cal S} ({\bm k}, {\bm x}_0) = \mathop {\lim }\limits_{N \to \infty} \frac{1}{N} \ln {\int {\prod\limits_{a = 1}^n {{\rm d}{{\bm{x}}_a}{{\rm e}^{ - \beta {{\left\| {{{\bm{x}}_a}} \right\|}_1}}}} \delta \left( {\bm{h}}\left( \left\{ {{{\bm{x}}_a}} \right\}_{a = 1}^n,{{\bm{x}}_0} \right) - {\bm{k}} \right)} }.
    	    \end{equation}
    	    
    	    We first express the constraint provided by RS ansatz as follow 
    	    \begin{eqnarray}
    	        && \delta \left( {\bm{h}}\left( \left\{ {{{\bm{x}}_a}} \right\}_{a = 1}^n,{{\bm{x}}_0} \right) - {\bm{k}} \right) \nonumber \\
    	        &\buildrel \textstyle. \over =& \frac{1}{\left(2 N \right)^{n^2 -1}} \prod\limits_{a = 1}^n {\left[ {\delta \left( {{{\left\| {{{\bm{x}}_a}} \right\|}_2^2} - 2NQ} \right)\delta \left( { {\rm{Re}} \left( { \bm{x}}_0^H {\bm{x}}_a \right) - 2Nm} \right)} \right]} \prod\limits_{a \ne b} {\delta \left( {{\bm{x}}_a^H {{\bm{x}}_b} - 2Nq} \right)}. 
    	    \end{eqnarray}
    	    We proceed with the derivation of \eqref{eq:ROMdu} by the Fourier transform of the Dirac's delta function:
            \begin{equation}
                \delta \left( {\bm x}_a^H  {\bm x}_b - 2N Q_{ab} \right) = \frac{1}{4 \pi j} {\int_{c - j \infty}^{c + j \infty} {\rm{d}} {\tilde Q_{ab}} \exp \left\{ \frac{\tilde Q_{ab}}{2} \left(  {\bm x}_a^H  {\bm x}_b - 2 N Q_{ab} \right) \right\} },
            \end{equation}
            where $c \in {\mathbb{R}}$ is an arbitrary real number.
            Therefore,
            \begin{eqnarray}
                && \int {\prod\limits_{a = 1}^n {{\rm d}{{\bm{x}}_a} {{\rm e}^{ - \beta {{\left\| {{{\bm{x}}_a}} \right\|}_1}}}} \delta \left( {\bm{h}}\left( \left\{ {{{\bm{x}}_a}} \right\}_{a = 1}^n,{{\bm{x}}_0} \right) - {\bm{k}} \right)} \nonumber \\
                &=& \int \prod\limits_{a, b} {\rm{d}} \frac{\tilde Q_{ab}}{4 \pi j} \int \prod\limits_{a = 1}^n {\rm d} {\bm{x}_a} \frac{1}{\left(2 N \right)^{n^2 -1}} \prod\limits_{a, b} \left( {\rm e}^{\frac{\tilde Q_{ab}}{2} {\bm x}_a^H  {\bm x}_b - N {\tilde Q_{ab}} Q_{ab} } \right) \prod\limits_{a = 1}^n { {{\rm e}^{ - \beta {{\left\| {{{\bm{x}}_a}} \right\|}_1}}}} \nonumber \\
                &=& \int \prod\limits_{a, b} {\rm{d}} \frac{\tilde Q_{ab}}{4 \pi j} \exp \left\{ N \left[ - \sum\limits_{a, b} {\tilde Q_{ab}} Q_{ab} + \frac{1}{N} \ln \int \prod\limits_{a = 1}^n {\rm d} {\bm{x}_a} \exp \left( \sum\limits_{a, b} \frac{\tilde Q_{ab}}{2} {\bm x}_a^H  {\bm x}_b - \sum\limits_{a = 1}^n \beta {{\left\| {{{\bm{x}}_a}} \right\|}_1} \right) \right] \right\}
            \end{eqnarray}
            With the saddle point method, we can approximately calculate the integral of $\tilde Q_{ab}$ by the function value of the saddle point $\tilde Q_{ab}^0$.
    	    Under the RS ansatz, we find that the saddle point $\tilde Q_{ab}^0$ have only three types of values, which is given as follow
    	    \begin{equation}
        	\label{eq:tildeQab}
        	    {\tilde{Q}_{ab}^0} = \left\{ {\begin{array}{*{20}{l}}
                {\begin{array}{*{20}{l}}
                {-\tilde{Q},}\\
                {\tilde{q},}\\
                {2 \tilde{m}.}
                \end{array}}&{\begin{array}{*{20}{l}}
                {(a = b = 1,2, \ldots ,n)}\\
                {(a > b = 1,2, \ldots ,n;b > a = 1,2, \ldots ,n)}\\
                {(a = 1,2, \ldots ,n,b = 0;a = 0,b = 1,2, \ldots ,n)}
                \end{array}}
                \end{array}} \right.
        	\end{equation}
        	Therefore,
        	\begin{eqnarray}
                &&{\cal S} ({\bm k}, {\bm x}_0) \nonumber \\
                &\buildrel \textstyle. \over =&{\cal S}(Q, q, m) \nonumber \\
                &=& \mathop {\rm{extr}}\limits_{\tilde Q,\tilde q,\tilde m} \left(  n \tilde Q Q - n (n - 1) \tilde q q - 2 n \tilde m m + \mathop {\lim }\limits_{N \to \infty} \frac{1}{N} {\cal R} \left( \tilde Q,\tilde q,\tilde m \right) \right),
        	\end{eqnarray}
        	where
        	\begin{eqnarray}
        	    && {\cal R} \left( \tilde Q,\tilde q,\tilde m \right) \nonumber \\
        	    &\buildrel \textstyle. \over = &  \ln \int \prod\limits_{a = 1}^n {\rm d} {\bm{x}_a} \exp \left( \sum\limits_{a = 1}^n -\frac{\tilde Q}{2} {\bm x}_a^H {\bm x}_a + \sum\limits_{a \ne b} \frac{\tilde q}{2} \bm{x}_a^H \bm{x}_b + \sum\limits_{a = 1}^n {\tilde m} {\rm{Re}} \left( \bm{x}_0^H \bm{x}_a \right) - \sum\limits_{a = 1}^n \beta \lambda \left\| \bm{x}_a \right\|_1 \right) \nonumber \\
        	    &=&  \sum\limits_{i = 1}^N \ln \int \prod\limits_{a = 1}^n {\rm d} x_{a, i} \exp \left( \sum\limits_{a = 1}^n -\frac{\tilde Q}{2} \left| x_{a, i} \right|^2 + \sum\limits_{a \ne b} \frac{\tilde q}{2} x_{a, i}^* x_{b, i} + \sum\limits_{a = 1}^n {\tilde m} {\rm{Re}} \left( x_{0, i}^* x_{a, i} \right) - \sum\limits_{a = 1}^n \beta \lambda \left| x_{a, i} \right| \right).
        	\end{eqnarray}
        	According to the well-known identity
        	\begin{equation}
        	    {\mathbb{E}}_z \exp \left( {\rm{Re}} \left(\sqrt{2}\lambda^* z \right) \right) = \exp{\left( \frac{\left| \lambda \right|^2}{2} \right)},
        	\end{equation}
        	where $z$ is a standard complex Gaussian variable, one can find that
        	\begin{eqnarray}
        	    && \exp \left( \sum\limits_{a \ne b} \frac{\tilde q}{2} x_{a, i}^* x_{b, i} \right) \nonumber \\
        	    &=& \exp{ \left( \frac{1}{2} \left| \sum\limits_{a = 1}^n {\sqrt{\tilde q} {x_{a, i}}} \right|^2 - \frac{\tilde q}{2}  \sum\limits_{a = 1}^n \left| x_{a, i} \right|^2 \right)}  \nonumber \\
        	    &=& \int {\rm{D}} {z_i} \exp \left( {\rm{Re}} \left( \sqrt{2 \tilde q} z_i^* \sum\limits_{a = 1}^n x_{a, i} \right) - \frac{\tilde q}{2} \sum\limits_{a = 1}^n \left| x_{a, i} \right|^2 \right).
        	\end{eqnarray}
        	Thus, 
        	\begin{eqnarray}
        	    && {\cal R} \left( \tilde Q,\tilde q,\tilde m \right) \nonumber \\
        	    &=& \sum\limits_{i = 1}^N \ln \int \prod\limits_{a = 1}^n {\rm d} x_{a, i} {\rm D} z_i \exp \left\{ \sum\limits_{a = 1}^n \left[ -\frac{\tilde Q + \tilde q}{2} \left| x_{a, i} \right|^2 + {\rm{Re}} \left( \left( \sqrt{2 \tilde q} z_i + \tilde m x_{0, i}  \right)^* x_{a, i} \right) - \beta \lambda \left| x_{a, i} \right| \right] \right\} \nonumber \\
        	    &=& \sum\limits_{i = 1}^N \ln \int {\rm D} z_i \left( \int {\rm d} x_i \exp \left\{ -\frac{\tilde Q + \tilde q}{2} \left| x_i \right|^2 + {\rm{Re}} \left( \left( \sqrt{2 \tilde q} z_i + \tilde m x_{0, i}  \right)^* x_i \right) - \beta \lambda \left| x_i \right|  \right\} \right)^n.
        	\end{eqnarray}
        	
    	\subsection{Final result}
    	\label{subsec:final}
    	    
    	    We continue working on the result of equation \eqref{eq:T+S}.
    	    Note that such result holds for $n = 1, 2, \ldots$.
        	In the following derivation, we assume the analytic continuation to real $n$ from the expression obtained by evaluating the relevant quantity only for positive integers $n$.
        	Thus,
        	\begin{eqnarray}
        	    &&\mathop {\lim }\limits_{n \to 0 } \frac{\partial }{{\partial n}} \mathop {\lim }\limits_{N \to \infty }  \frac{1}{N} \ln {\mathbb{E}}_{{\bm A}, {\bm \xi}} \left[ Z^n \right] \nonumber \\
        	    &=& \mathop {\lim }\limits_{n \to 0 } \frac{\partial }{{\partial n}} \mathop {\rm{extr}}\limits_{Q, q, m}  \left\{ {{\cal T}}\left( {Q,q,m} \right) + {{\cal S}}\left( {Q,q,m} \right) \right\} \nonumber \\
        	    &=& \mathop {\rm{extr}}\limits_{Q, q, m}  \left\{ \mathop {\lim }\limits_{n \to 0 } \frac{\partial }{{\partial n}} {{\cal T}}\left( {Q,q,m} \right) + \mathop {\lim }\limits_{n \to 0 } \frac{\partial }{{\partial n}} {{\cal S}}\left( {Q,q,m} \right) \right\},
        	\end{eqnarray}
    	    where 
    	    \begin{eqnarray}
    	        &&\mathop {\lim }\limits_{n \to 0 } \frac{\partial }{{\partial n}} {{\cal T}}\left( {Q,q,m} \right) \nonumber \\
    	        &=& \mathop {\lim }\limits_{n \to 0 } \frac{\partial }{{\partial n}} \left\{ G \left( - \frac{2\beta \left( Q - q + n(q - 2m + \rho) \right)}{2 + \beta n \sigma^2}; {\bm J} \right) \right. \nonumber \\
                && \left. + (n-1) G(-\beta(Q-q); {\bm J}) - \frac{\gamma}{2} \ln \left( 1 + \beta n \sigma^2 \right) \right\} \nonumber \\
    	        &=& -G' \left( -\beta \left( Q - q \right) ; {\bm J}  \right) \left( \beta \left( q - 2m + \rho \right) - \frac{\beta^2 \sigma^2 \left( Q - q \right)}{2} \right) \nonumber \\
    	        && + G \left( -\beta \left( Q - q \right); {\bm J} \right) - \frac{\beta \gamma \sigma^2}{2},
    	    \end{eqnarray}
    	    and
    	    \begin{eqnarray}
    	        &&\mathop {\lim }\limits_{n \to 0 } \frac{\partial }{{\partial n}} {{\cal S}}\left( {Q,q,m} \right) \nonumber \\
    	        &=& \mathop {\rm{extr}}\limits_{\tilde Q, \tilde q, \tilde m}  \left\{ \mathop {\lim }\limits_{n \to 0 } \frac{\partial }{{\partial n}} \left[ n \tilde Q Q - n (n - 1) \tilde q q - 2 n \tilde m m + \mathop {\lim }\limits_{N \to \infty} \frac{1}{N} {\cal R} \left( \tilde Q,\tilde q,\tilde m \right) \right] \right\}  \nonumber \\
    	        &=& \mathop {\rm{extr}}\limits_{\tilde Q, \tilde q, \tilde m}  \left\{ \tilde Q Q + \tilde q q - 2 \tilde m m + \mathop {\lim }\limits_{N \to \infty} \frac{1}{N} \mathop {\lim }\limits_{n \to 0 } \frac{\partial }{{\partial n}} {\cal R} \left( \tilde Q,\tilde q,\tilde m \right)  \right\}.
    	    \end{eqnarray}
    	    The part of the partial of ${\cal R} \left( \tilde Q,\tilde q,\tilde m \right)$ is given by 
    	    \begin{eqnarray}
    	        && \mathop {\lim }\limits_{n \to 0 } \frac{\partial }{{\partial n}} {\cal R} \left( \tilde Q,\tilde q,\tilde m \right) \nonumber \\
    	        &=& \mathop {\lim }\limits_{n \to 0 } \frac{\partial }{{\partial n}} \sum\limits_{i = 1}^N \ln \int {\rm D} z_i \left( \int {\rm d} x_i \exp \left\{ -\frac{\tilde Q + \tilde q}{2} \left| x_i \right|^2 + {\rm{Re}} \left( \left( \sqrt{2 \tilde q} z_i + \tilde m x_{0, i}  \right)^* x_i \right) - \beta \lambda \left| x_i \right|  \right\} \right)^n \nonumber \\
    	        &=& \sum\limits_{i = 1}^N \int {\rm D} z_i \ln \left( \int {\rm d} x_i \exp \left\{ -\frac{\tilde Q + \tilde q}{2} \left| x_i \right|^2 + {\rm{Re}} \left( \left( \sqrt{2 \tilde q} z_i + \tilde m x_{0, i}  \right)^* x_i \right) - \beta \lambda \left| x_i \right|  \right\} \right).
    	    \end{eqnarray}
    	    
    	    When $\beta \to \infty$, the following substitutions of variables are employed
    	    \begin{eqnarray*}
    	        \chi &\buildrel \textstyle. \over =& \beta (Q - q), \\
    	        \hat Q &\buildrel \textstyle. \over =& \beta^{-1} (\tilde Q + \tilde q), \\
    	        \hat \chi &\buildrel \textstyle. \over =& \beta^{-2} \tilde q, \\
    	        \hat m &\buildrel \textstyle. \over =& \beta^{-1} \tilde m.
    	    \end{eqnarray*}
    	    With these substitutions, we have
    	    \begin{eqnarray}
    	        \label{eq:beta_n_T}
    	        && \mathop {\lim }\limits_{\beta \to \infty } -\frac{1}{\beta} \mathop {\lim }\limits_{n \to 0 } \frac{\partial }{{\partial n}} {{\cal T}}\left( {Q,q,m} \right) \nonumber \\
    	        &=& \mathop {\lim }\limits_{\beta \to \infty } -\frac{1}{\beta} \left[ -G' \left( -\beta \left( Q - q \right) ; {\bm J} \right) \left( \beta \left( q - 2m + \rho \right) - \frac{\beta^2 \sigma^2 \left( Q - q \right)}{2} \right) \right. \nonumber \\
    	        && \left. + G \left( -\beta \left( Q - q \right); {\bm J} \right) - \frac{\beta \gamma \sigma^2}{2} \right] \nonumber \\
    	        &=& G' (-\chi; {\bm{J}}) (Q - 2m + \rho - \frac{\chi}{2} \sigma^2) + \frac{\gamma}{2}\sigma^2,
    	    \end{eqnarray}
    	    and
    	    \begin{eqnarray}
    	        \label{eq:beta_n_S}
    	        && \mathop {\lim }\limits_{\beta \to \infty } -\frac{1}{\beta} \mathop {\lim }\limits_{n \to 0 } \frac{\partial }{{\partial n}} {{\cal S}_n}\left( {Q,q,m} \right) \nonumber \\
    	        &=& \mathop {\rm{extr}}\limits_{\tilde Q, \tilde q, \tilde m}  \left\{ \mathop {\lim }\limits_{\beta \to \infty } -\frac{1}{\beta} \left[  \tilde Q Q + \tilde q q - 2 \tilde m m + \mathop {\lim }\limits_{N \to \infty} \frac{1}{N} \mathop {\lim }\limits_{n \to 0 } \frac{\partial }{{\partial n}} {\cal R} \left( \tilde Q,\tilde q,\tilde m \right) \right] \right\}  \nonumber \\
    	        &=& \mathop {\rm{extr}}\limits_{\tilde Q, \tilde q, \tilde m}  \left\{ - {\hat Q Q} + {\hat \chi \chi} + 2 \hat m m  + \mathop {\lim }\limits_{\beta \to \infty } -\frac{1}{\beta} \mathop {\lim }\limits_{N \to \infty} \frac{1}{N} \mathop {\lim }\limits_{n \to 0 } \frac{\partial }{{\partial n}} {\cal R} \left( \tilde Q,\tilde q,\tilde m \right)  \right\}.
    	    \end{eqnarray}
    	    With the saddle point method, one has
    	    \begin{eqnarray}
    	        \label{eq:beta_n_R}
    	        && \mathop {\lim }\limits_{\beta \to \infty } -\frac{1}{\beta} \mathop {\lim }\limits_{N \to \infty} \frac{1}{N} \mathop {\lim }\limits_{n \to 0 } \frac{\partial }{{\partial n}} {\cal R} \left( \tilde Q,\tilde q,\tilde m \right) \nonumber \\
    	        &=& \mathop {\lim }\limits_{N \to \infty} \frac{1}{N} \mathop {\lim }\limits_{\beta \to \infty } -\frac{1}{\beta} \sum\limits_{i = 1}^N \int {\rm D} z_i \ln \left( \int {\rm d} x_i \exp \left\{ -\frac{\tilde Q + \tilde q}{2} \left| x_i \right|^2 + {\rm{Re}} \left( \left( \sqrt{2 \tilde q} z_i + \tilde m x_{0, i}  \right)^* x_i \right) - \beta \lambda \left| x_i \right|  \right\} \right) \nonumber \\
    	        &=& \mathop {\lim }\limits_{N \to \infty} \frac{1}{N} \sum\limits_{i = 1}^N \int {\rm D} z_i \mathop {\lim }\limits_{\beta \to \infty } -\frac{1}{\beta} \ln \left( \int {\rm d} x_i \exp \left\{ - \beta \left[ \frac{\hat Q}{2} \left| x_i \right|^2 - {\rm{Re}} \left( \left( \sqrt{2 \hat \chi} z_i + \hat m x_{0, i}  \right)^* x_i \right) + \lambda \left| x_i \right| \right] \right\} \right) \nonumber \\
    	        &=& \mathop {\lim }\limits_{N \to \infty } \frac{1}{N} \sum\limits_{i = 1}^{N} {\int {\mathop {\min }\limits_{{x_i}} \left[ { \frac{{\hat Q}}{2}\left| {x_i} \right|^2 - {\rm{Re}} \left( \left( {\hat m{x_{0,i}} + \sqrt {2 \hat \chi } {z_i}} \right)^*{x_i} \right) + \lambda \left| {x_i} \right|} \right]{\rm{D}}{z_i}} }.
    	    \end{eqnarray}
    	    
    	    Finally, associating with the result \eqref{eq:beta_n_T}, \eqref{eq:beta_n_S} and \eqref{eq:beta_n_R} yields the expression of the average free energy density in zero-temperature limit
    	    \begin{eqnarray}
                f &=& \mathop {\rm{extr}} \limits_{ Q, \hat Q, \chi, \hat \chi, m, \hat m } \left\{ G' (-\chi; {\bm{J}}) (Q - 2m + \rho - \frac{\chi}{2} \sigma^2) + \frac{\gamma}{2}\sigma^2 - {\hat QQ} + {\hat \chi \chi} + 2 \hat mm \right. \nonumber \\ 
                && \left. + \mathop {\lim }\limits_{N \to \infty } \frac{1}{N} \sum\limits_{i = 1}^{N} {\int {\mathop {\min }\limits_{{x_i}} \left[ { \frac{{\hat Q}}{2}\left| {x_i} \right|^2 - {\rm{Re}} \left( \left( {\hat m{x_{0,i}} + \sqrt {2 \hat \chi } {z_i}} \right)^*{x_i} \right) + \lambda \left| {x_i} \right|} \right]{\rm{D}}{z_i}} }  \right\}.
            \end{eqnarray}
    
    \section{Derivation of averages with respect to row-orthogonal matrices}
    \label{appendix:ROM}
        
        We here would like to derive the averages with respect to row-orthogonal matrices ${\bm A} \in {\mathbb C}^{M \times N}$ as follow:
        \begin{equation}
        \label{eq:ROMdA}
            L_{\rm ROM}(-{\bm x}) \buildrel \textstyle. \over = \mathop {\lim }\limits_{N \to \infty} \frac{1}{N} \ln {\mathbb E}_{\bm A}\left[ \exp \left\{ - \sum\limits_{i = 1}^n \frac{1}{2} {\bm v}_i^H {\bm A}^H {\bm A} {\bm v}_i \right\} \right],
        \end{equation}
        where ${\bm v}_i \in {\mathbb C}^{N \times 1}$ $(1 \le i \le n)$ is a set of orthogonal vectors satisfying ${\bm v}_i^H {\bm v}_j = 2N \nu_i \delta_{ij}$, $\delta_{i j}$ is unity for $i = j$ and vanishes otherwise, ${\bm x} = \left[ \nu_1, \nu_2, \ldots,  \nu_n \right]^T$ and the average is taken over all the row-orthogonal matrices.
        Denote by ${\bm J} = {\bm A}^H {\bm A}$, then the Hermitian matrix ${\bm J} \in {\mathbb C}^{N \times N}$ can be factorized as ${\bm J} = {\bm O}{\bm D}{\bm O}^H$ with an orthogonal matrix ${\bm O} \in {\mathbb C}^{N \times N}$ and a diagonal matrix ${\bm D}$, in which its on-diagonal elements are eigenvalues of ${\bm J}$.
        Due to the fact that ${\bm A}$ is row-orthogonal, the eigenvalues of ${\bm J}$ are fixed to $1$ or $0$, that is, ${\bm D}$ does not vary with ${\bm A}$.
        In addition, denote by $\tilde {\bm v}_i = {\bm O}^H {\bm v}_i$, the $\ell_2$-norm of $\tilde {\bm v}_i$ is maintained: $\tilde {\bm v}_i^H \tilde {\bm v}_i = 2N \nu_i$.
        Therefore, the average on ${\bm A}$ reduces to the average on $\tilde {\bm v}_i$, which is given by
        \begin{equation}
        \label{eq:ROMdu}
            L_{\rm ROM}(-{\bm x}) = \mathop {\lim }\limits_{N \to \infty} \frac{1}{N} \ln \left\{ \frac{ {\int {\prod \limits_{i = 1}^n {\rm{d}} \tilde {\bm v}_i } \exp{\left( - \sum\limits_{i = 1}^n \frac{1}{2} \tilde {\bm v}_i^H {\bm D} \tilde {\bm v}_i \right)} {\prod \limits_{i = 1}^n \delta \left( \tilde {\bm v}_i^H \tilde {\bm v}_i - 2N \nu_i \right) } } }{ {\int {\prod \limits_{i = 1}^n {\rm{d}} \tilde {\bm v}_i } {\prod \limits_{i = 1}^n \delta \left( \tilde {\bm v}_i^H \tilde {\bm v}_i - 2N \nu_i \right) } } } \right\}.
        \end{equation}
        Note that here we discard the fact that $\left\{ {\bm v}_i \right\}_{i=1}^n$ are orthogonal to each other, for when the dimension of vectors $N \to \infty$, the probability that any finite number of vectors are orthogonal to each other is $1$.
        
        We proceed with the derivation of \eqref{eq:ROMdu} by the Fourier transform of the Dirac's delta function:
        \begin{equation}
            \delta \left( \tilde {\bm v}_i^H \tilde {\bm v}_i - 2N \nu_i \right) = \frac{1}{4 \pi j} {\int_{c - j \infty}^{c + j \infty} {\rm{d}}\Lambda_i \exp \left\{ - \frac{\Lambda_i}{2} \left( \tilde {\bm v}_i^H \tilde {\bm v}_i - 2 N \nu_i \right) \right\} },
        \end{equation}
        where $c \in {\mathbb{R}}$ is an arbitrary real number.
        Thus, 
        \begin{equation}
        \label{eq:ROMdu_denominator}
            {\int {\prod \limits_{i = 1}^n \left[ {\rm{d}} \tilde {\bm v}_i  \delta \left( \tilde {\bm v}_i^H \tilde {\bm v}_i - 2N \nu_i \right) \right]}} = \frac{\left( 2 \pi \right)^{n N}}{\left(4 \pi j \right)^n} {\int_{c - j \infty}^{c + j \infty} {\prod \limits_{i = 1}^n {\rm{d}}\Lambda_i} \exp \left\{ - N  {\sum \limits_{i = 1}^n \left(\ln \Lambda_i -  \Lambda_i \nu_i \right) } \right\} },
        \end{equation}
        where we use the complex Gaussian integration formula
        \begin{equation}
            \frac{1}{\left( 2 \pi \right)^N} {\int \exp{ \left( -\frac{1}{2} {\bm z}^H {\bm M} {\bm z} + {\rm{Re}} \left( {\bm b}^H {\bm z} \right) \right) } {\rm{d}} {\bm z} } = \frac{1}{\det {\bm M} } \exp{ \left( \frac{1}{2} {\bm b}^H {\bm M}^{-1} {\bm b} \right) },
        \end{equation}
        where ${\bm b}, {\bm z} \in {\mathbb{C}}^{N}$ and ${\bm M}$ is symmetric positive definite.

        Applying saddle-point method, we rewrite \eqref{eq:ROMdu_denominator} as
        \begin{eqnarray}
            \label{eq:ROMdu_denominator2}
            &&{\int {\prod \limits_{i = 1}^n \left[ {\rm{d}} \tilde {\bm v}_i  \delta \left( \tilde {\bm v}_i^H \tilde {\bm v}_i - 2N \nu_i \right) \right]}} \nonumber \\
            &=& \left( 2 \pi \right)^{n N} \exp \left\{ -N { \left( \sum \limits_{i = 1}^n  \mathop {{\rm{extr}}} \limits_{\left\{ \Lambda_i \right\}} \left[  \ln \Lambda_i -  \Lambda_i \nu_i \right] +  O \left( \frac{1}{N} \right) \right) } \right\}  \nonumber \\
            &=& \left( 2 \pi \right)^{n N} \exp \left\{ -N { \left( \sum \limits_{i = 1}^n \left( -  \ln \nu_i - 1 \right) +  O \left( \frac{1}{N} \right) \right) } \right\}.
        \end{eqnarray}
        With similar procedure, the numerator of \eqref{eq:ROMdu} can be expressed as
        \begin{eqnarray}
        \label{eq:ROMdu_numerator}
            &&{\int {\prod \limits_{i = 1}^n {\rm{d}} \tilde {\bm v}_i } \exp{\left( - \sum\limits_{i = 1}^n \frac{1}{2} \tilde {\bm v}_i^H {\bm D} \tilde {\bm v}_i \right)} {\prod \limits_{i = 1}^n \delta \left( \tilde {\bm v}_i^H \tilde {\bm v}_i - 2N \nu_i \right) } } \nonumber \\
            &=& \frac{1}{\left(4 \pi j \right)^n} {\int_{c - j \infty}^{c + j \infty} {\prod \limits_{i = 1}^n {\rm{d}}\Lambda_i} \int {\prod \limits_{i = 1}^n {\rm{d}} \tilde {\bm v}_i } \exp \left\{  \sum\limits_{i = 1}^n \left( -\frac{1}{2} \tilde {\bm v}_i^H \left( \Lambda_i {\bm I}_N + {\bm D} \right) \tilde {\bm v}_i + N \Lambda_i \nu_i \right) \right\} } \nonumber \\
            &=& \frac{\left( 2 \pi \right)^{n N}}{\left(4 \pi j \right)^n} {\int_{c - j \infty}^{c + j \infty} {\prod \limits_{i = 1}^n  {\rm{d}}\Lambda_i} \exp \left\{  \sum\limits_{i = 1}^n \left( -\sum\limits_{m = 1}^N \ln \left( \Lambda_i + d_m \right) + N \Lambda_i \nu_i \right) \right\} } \nonumber \\
            &=& \frac{\left( 2 \pi \right)^{n N}}{\left(4 \pi j \right)^n} {\int_{c - j \infty}^{c + j \infty} {\prod \limits_{i = 1}^n {\rm{d}}\Lambda_i} \exp \left\{ N \left( \sum\limits_{i = 1}^n \left( - \frac{1}{N} \sum\limits_{m = 1}^N \ln \left( \Lambda_i + d_m \right) + \Lambda_i \nu_i \right) \right) \right\} },
        \end{eqnarray}
        where $d_m$ $(1 \le m \le N)$ are the diagonal elements of ${\bm D}$, thus are the eigenvalues of ${\bm J}$.
        When $N$ are large enough, we have the following approximation:
        \begin{equation}
            \frac{1}{N} \sum\limits_{m = 1}^N \ln \left( \Lambda_i + d_m \right) = \int {\rm{d}} \lambda \rho_{\bm J} (\lambda) \ln \left( \Lambda_i + \lambda \right).
        \end{equation}
        Therefore, by using such approximation and combining \eqref{eq:ROMdu_denominator2} and \eqref{eq:ROMdu_numerator}, \eqref{eq:ROMdu} can be expressed as
        \small
        \begin{eqnarray}
            L_{\rm ROM}(-{\bm x}) &=& \mathop {\lim }\limits_{N \to \infty} \frac{1}{N} \ln \left\{  \frac{1}{\left(4 \pi j \right)^n} {\int_{c - j \infty}^{c + j \infty} {\prod \limits_{i = 1}^n  {\rm{d}}\Lambda_i} \cdot } \right. \nonumber \\
            && \left. \exp \left[ N \left( \sum\limits_{i = 1}^n \left( - \int {\rm{d}} \lambda \rho_{\bm J} (\lambda) \ln \left( \Lambda_i + \lambda \right) + \Lambda_i \nu_i - \ln \nu_i - 1 \right) + O \left( \frac{1}{N} \right) \right) \right]  \right\} \nonumber \\
            &=& \sum\limits_{i = 1}^n \left( \mathop {\rm{extr} }\limits_{\Lambda_i} \left[ - \int {\rm{d}} \lambda \rho_{\bm J} (\lambda) \ln \left( \Lambda_i + \lambda \right) + \Lambda_i  \nu_i \right] - \ln  \nu_i - 1 \right)\nonumber \\
            &=& \sum\limits_{i = 1}^n \left( \mathop {\rm{extr} }\limits_{\Lambda_i} \left[ - \int {\rm{d}} \lambda \rho_{\bm J} (\lambda) \ln \left| \Lambda_i - \lambda \right| - \Lambda_i  \nu_i \right] - \ln \left| -  \nu_i \right| - 1 \right).
        \end{eqnarray}
        \normalsize
        Recall the definition of function $G$:
        \begin{equation}
            G(x; {\bm{J}}) \buildrel \textstyle. \over = \mathop {\rm{extr}} \limits_{z} \left[ -\int{\rho_{\bm J}(s) \ln \left| z - s \right| {\rm{d}}s} + zx \right] - \ln \left| x \right| - 1.
        \end{equation}
        We finally obtain:
        \begin{equation}
            L_{\rm ROM}(-{\bm x}) = \sum\limits_{i = 1}^n G(-x_i; {\bm{J}}).
        \end{equation}
        
    \section{Derivation of $\chi$ through linear response argument}
    \label{appendix:LRA}
        
        Denote by 
        \begin{equation}
            Gi_0({\bm m}, {\bm h}, Q, \Lambda) = {\rm{Re}} \left( {\bm h}^H {\bm m} \right) - N \Lambda Q - \frac{1}{\beta} \ln \int{ {\rm{e}}^{ - \frac{\beta}{2} \left\| {\bm y} - {\bm A} {\bm x} \right\|_2^2 + \beta {\rm{Re}} \left( {\bm h}^H {\bm x} \right) -\frac{\beta}{2} \Lambda \left\| {\bm x} \right\|_2^2 - \beta \lambda \left\| {\bm x} \right\|_1 } {\rm{d}}{\bm x} },
        \end{equation}
        By recalling the definition of the Boltzmann average $\left \langle {\bm x} \right \rangle$, one can find that
        \begin{equation}
             \left. \frac{\partial Gi_0}{\partial {\rm{Re}} \left(h_i \right)} \right|_{h_i = h_{0, i}} = {\rm{Re}} \left(m_i \right) - {\rm{Re}} \left( \left \langle x_i \right \rangle \right),
,        \end{equation}
        \begin{equation}
             \left. \frac{\partial Gi_0}{\partial {\rm{Im}} \left(h_i \right)} \right|_{h_i = h_{0, i}} = {\rm{Im}} \left(m_i \right) - {\rm{Im}} \left( \left \langle x_i \right \rangle \right),
        \end{equation}
        and further
        \begin{equation}
             \left. \frac{\partial^2 Gi_0}{\partial {\rm{Re}} \left(h_i \right)^2} \right|_{h_i = h_{0, i}} + \left. \frac{\partial^2 Gi_0}{\partial {\rm{Im}} \left(h_i \right)^2} \right|_{h_i = h_{0, i}} = \beta \left( \left \langle \left| x_i \right|^2 \right \rangle - \left| \left \langle x_i \right \rangle \right|^2 \right),
        \end{equation}
        where ${\bm h}_{0}$ is the extreme value of $G_0$, $h_{i}$ and $h_{0, i}$ is the $i$-th entry of ${\bm h}$ and ${\bm h}_{0}$, respectively. 
        Therefore,
        \begin{equation}
            \chi = \frac{1}{2N} {\sum \limits_{i = 1}^N \left( \left. \frac{\partial^2 Gi_0}{\partial {\rm{Re}} \left(h_i \right)^2} \right|_{h_i = h_{0, i}} + \left. \frac{\partial^2 Gi_0}{\partial {\rm{Im}} \left(h_i \right)^2} \right|_{h_i = h_{0, i}} \right) }.
        \end{equation}
        We compute $\chi$ with the following approximation: \begin{equation}
        \label{eq:chiapprox}
             \chi = \frac{1}{2N} {\sum \limits_{i = 1}^N \left( \left. \frac{\partial^2 Gi_1}{\partial {\rm{Re}} \left(h_i \right)^2} \right|_{h_i = h_{1, i}} + \left. \frac{\partial^2 Gi_1}{\partial {\rm{Im}} \left(h_i \right)^2} \right|_{h_i = h_{1, i}} \right) },
        \end{equation}
        where 
        \begin{eqnarray}
            Gi_1({\bm m}, {\bm h}, Q, \Lambda) &=&  {\rm{Re}} \left( {\bm h}^H {\bm m} \right) - N \Lambda Q - {\sum \limits_{i = 1}^N \frac{\left( \left| h_i \right| - \lambda \right)^2}{2 \Lambda} \cdot \Theta \left( \left| h_i \right| - \lambda \right) } \nonumber \\
	        &&- \frac{N}{\beta} G(-\chi; {\bm{J}}) + \frac{1}{2} \left\| {\bm y} - {\bm A} {\bm m} \right\|_2^2.
        \end{eqnarray}
        and ${\bm h}_{1}$ is the extreme value of $Gi_1$ with $h_{1, i}$ denoting the $i$-th entry of ${\bm h}_{1}$.
        Hence, \eqref{eq:getchi} can be derived from \eqref{eq:chiapprox}.

	\bibliographystyle{IEEEtran}
	\bibliography{IEEEabrv,references}
					
\end{document}